\documentclass[twocolumn,showpacs,aps,prd,floatfix]{revtex4}
\usepackage{graphicx}
\usepackage{dcolumn}
\usepackage{amsmath}
\usepackage{epsfig}
\usepackage{multirow}
\usepackage{color}
\usepackage{bm}

\input pubboard/babarsym

\newcommand{\BABARPubYear}    {06}
\newcommand{\BABARPubNumber}  {020}
\newcommand{\SLACPubNumber} {11898}

\def\Dsp      {\ensuremath{D^+_s}\xspace}
\def\Dsm      {\ensuremath{D^-_s}\xspace}
\def\Ds       {\ensuremath{D_s}\xspace}
\def\Dspm     {\ensuremath{D^\pm_s}\xspace}

\def\Lambdabar {\kern 0.2em\overline{\kern -0.2em \Lambda}{}\xspace}
\def\fbar {\ensuremath{\kern 0.2em\overline{\kern -0.2em f}{}}\xspace}
\def\XIbar {\kern 0.2em\overline{\kern -0.2em \Xi}{}\xspace}
\def\BRbar {\kern 0.2em\overline{\kern -0.2em \BR}{}\xspace}

\def\Lcp {\ensuremath{\Lambda^+_c}\xspace}
\def\Lcm {\ensuremath{\Lambdabar^-_c}\xspace}
\def\Lc  {\ensuremath{\Lcp}\xspace}
\def\Lcpm{\ensuremath{\Lambda^{\pm}_c}\xspace}

\def\Xib  {\ensuremath{\XIbar}\xspace}

\def\Xic  {\ensuremath{\Xi_c}\xspace}
\def\Xicb  {\ensuremath{\Xib_c}\xspace}

\def\dkpi      {\Dz\to\Km\pip}

\def\dkpipipi  {\Dz\to\Km\pip\pim\pip}

\def\dkpipi    {\Dp\to\Km\pip\pip}

\def\dsphipi   {\ensuremath{\Dsp\to\phi\pip}}
\def\dskstark  {\Dsp\to\Kstarzb\Kp}
\def\dsksk     {\Dsp\to\KS\Kp}
\def\lcpkpi    {\ensuremath{\Lcp\to\proton\,\Km\pip}}

\def\Brec  {\ensuremath{\B_{\mathrm{rec'd}}}\xspace}
\def\Brecp {\ensuremath{\Brec^+}\xspace}
\def\Brecz {\ensuremath{\Brec^0}\xspace}

\def\Cbar {\kern 0.2em\overline{\kern -0.2em C}{}\xspace}

\def\Xc  {\ensuremath{C}\xspace}
\def\Xbc {\ensuremath{\Cbar}\xspace}

\def\rt {\ensuremath{\frac{\tau_{\Bp}}{\tau_{\Bz}}}\xspace }

\def\Xccbar {\ensuremath{\Xc\,(\Xbc)}\xspace}

\def\ARGUS{\ensuremath{ARG}\xspace}

\begin{document}

\preprint{\babar-PUB-\BABARPubYear/\BABARPubNumber}
\preprint{SLAC-PUB-\SLACPubNumber}

\begin{flushleft}
\babar-PUB-\BABARPubYear/\BABARPubNumber\\
SLAC-PUB-\SLACPubNumber\\
\end{flushleft}

\title{
{\large \bf Study of Inclusive \boldmath \Bm and \Bzb Decays to Flavor-Tagged $D$, \Ds and \Lc } }

%
\author{B.~Aubert}
\author{R.~Barate}
\author{M.~Bona}
\author{D.~Boutigny}
\author{F.~Couderc}
\author{Y.~Karyotakis}
\author{J.~P.~Lees}
\author{V.~Poireau}
\author{V.~Tisserand}
\author{A.~Zghiche}
\affiliation{Laboratoire de Physique des Particules, F-74941 Annecy-le-Vieux, France }
% ...
\collaboration{The \babar\ Collaboration}
\noaffiliation

\date{\today}

\begin{abstract}
We report on a study of inclusive \Bm and \Bzb meson decays to ${\Dz \X}$, ${\Dzb \X}$, ${\Dp \X}$, ${\Dm \X}$, ${\Dsp \X}$, ${\Dsm \X}$, ${\Lcp \X}$, ${\Lcm \X}$, based on a sample of 231 million $\BB$ events recorded with the \babar\ detector at the \Y4S resonance. Events are selected by completely reconstructing one $\B$ and searching for a reconstructed charm particle in the rest of the event. From the measured branching fractions of these decays, we infer the number of charm and anti-charm particles per \Bb decay, separately for charged and neutral parents. We derive the total charm yield per \Bm decay, $n_\c^- = 1.202 \pm 0.023\pm 0.040^{+0.035}_{-0.029} $, and per \Bzb decay, $n_\c^0 = 1.193 \pm 0.030\pm 0.034^{+0.044}_{-0.035}$  where the first uncertainty is statistical, the second is systematic, and the third reflects the charm branching-fraction uncertainties. We also present the charm momentum distributions measured in the \Bb rest frame.
\end{abstract}


\maketitle

\section{Introduction}
\label{introduction}

The dominant process for the decay of a \b quark is $\b\to\c \W^{*-}$~\cite{chconj}, resulting in a (flavor) correlated \c quark and a virtual \W. In the decay of the \W, the production of a $\ubar d$ or a $\cbar s$ pair are both Cabibbo-allowed and should be approximately equal, the latter being suppressed by a phase-space factor. The first process dominates hadronic \b decays. The second can be easily distinguished as it produces a (flavor) anticorrelated \cbar quark. Experimentally, we investigate correlated and anticorrelated charm production through the measurement of the inclusive \B-decay rates to a limited number of charm hadron species, i.e. \Dz, \Dzb, \Dp, \Dm, \Dsp, \Dsm, \Lcp, \Lcm, \Xic and charmonia, because all other charm particles decay into one of the previous hadrons. 

The analysis presented here exploits a substantially larger data sample than the original \babar\ result~\cite{babarnc}. It also employs a more sophisticated fitting method to extract, in a correlated manner, the number of reconstructed \B mesons and the charm hadron yields, which reduces the experimental systematic uncertainty. Other measurements~\cite{cleonc,cleoinc,delphiinc,cleolc,belleinc} of these rates are more statistically limited and/or do not distinguish between the different parent~\B states. Besides the theoretical interest~\cite{ncbagan,ncbuchalla,ncneubert,nc_lenz}, the fact that anticorrelated charm particles are a background for many studies also motivates a more precise measurement of their production rates in \B decays.\\

Most of the charged and neutral $D$ mesons produced in \Bb decays come from correlated production $\Bb \to D X$. However, a significant number of $\Bb \to \Db X$ decays are expected through $\b\to\c\cbar\s$ transitions, such as $\Bb \to D^{(*)}\Db^{(*)} \Kb^{(*)}(n\pi)$. Although the branching fractions of the 3-body decays $\Bb\to D^{(*)} \Db^{(*)} \Kb$ have been measured~\cite{alephddk,babarddk}, they do not saturate $\Bb\to \Db X$ transitions~\cite{babarnc}. It is therefore important to improve the precision on the $\Bb\to \Db X$ branching fraction.

By contrast, anticorrelated \Dsm production, ${\Bb\to \Dsm D (n\pi)}$, is expected to dominate \Bb decays to \Ds mesons, since correlated production needs an extra \ssbar pair created from the vacuum to give ${\Bb\to \Dsp \Km (n\pi)}$. There is no prior published measurement for correlated \Dsp production.

Correlated \Lc are produced in decays like $\Bb\to\Lcp \antiproton \pim(\pi)$, while anticorrelated $\Lcm$ should originate predominantly from $\Bb\to\Xic\Lcm(\pi)$. The decay $\Bb\to\Xic\Lcm$ has recently been observed~\cite{belleXicLc}, confirming the hypothesis of associated \Xic\Lcm production. Another possibility for anticorrelated \Lcm production is $\Bb\to \Lcp \Lcm \kaon$, the baryonic analogue of the $D \Db \kaon$ decay.\\

This analysis uses \upsbb events in which either a \Bp or a \Bz meson  (hereafter denoted \Brec) decays into a hadronic final state and is fully reconstructed. We then reconstruct $D$, \Ds and \Lc from the decay products of the recoiling \Bm (\Bzb) meson and compare the flavor of the charm hadron with that of the reconstructed \B (taking into account \Bz-\Bzb mixing). This allows separate measurements of the \Bm\ (\Bzb)\ \to\Dz\X, \Dp\X, \Dsp\X, \Lcp\X and \Bm\ (\Bzb)\ \to\Dzb \X, \Dm\X, \Dsm\X, \Lcm\X branching fractions. 

We then compute the average number of correlated (anticorrelated) charm particles per \Bm decay, $N_\c^- $ ($N_{\cbar}^-$)~:
\begin{eqnarray}
    N_\c^-      & = & \sum_{\Xc}    \BR(\Bm\to \Xc X),\label{eq:nc} \\
    N_{\cbar}^- & = & \sum_{\Xbc} \BR(\Bm\to \Xbc X), \label{eq:ncbar}
\end{eqnarray}
where the sum is performed over $\Xc\equiv\ \{\Dz,\ \Dp,\ \Dsp,\ \Lcp,\ \Xic,\ (\ccbar)\}$ or $\Xbc\equiv\ \{\Dzb,\ \Dm,\ \Dsm,\ \Lcm,\ (\ccbar)\}$, where $(\ccbar)$ refers to all charmonium states collectively. We neglect anticorrelated \Xicb production, as it requires both a $\cbar\s$ and an \ssbar pair in the decay to give $\Xicb\Omega_c$. We then sum $N_\c^-$ and $N_{\cbar}^-$ to obtain the average number of charm plus anti-charm quarks per \Bm decay, $n_\c^- = N_\c^- + N_{\cbar}^-$. We similarly define $N_{\c}^0$, $N_{\cbar}^0$ and $n_c^0$ for \Bzb decays.\\

The above method also lends itself to a measurement of the momentum distribution of each charm species directly in the rest frame of the parent meson, because the four-momentum of each recoiling \Bb is fully determined from those of the \Y4S and of the reconstructed \B. The resulting charm spectra can then be compared to theoretical predictions in the same frame~\cite{bauer_dx}. This avoids the significant smearing due to the Lorentz boost from the parent-\Bb frame to the \Y4S frame affecting earlier measurements, such as those reported in~\cite{cleonc}. These spectra might also show indications of four-quark states~\cite{bigi_4q}. 

\section{\babar\ detector and data sample}

The measurements presented here are based on a sample of $231$ million \BB pairs ($210~\invfb$) recorded at the \Y4S resonance with the \babar\ detector at the \pep2 asymmetric-energy \B factory at SLAC. The \babar\ detector is described in detail elsewhere~\cite{det}. Charged-particle trajectories are measured by a 5-layer double-sided silicon vertex tracker and a 40-layer drift chamber, both operating in a 1.5-T solenoidal magnetic field. Charged-particle identification is provided by the average energy loss (\dedx) in the tracking devices and by an internally reflecting ring-imaging Cherenkov detector. Photons are detected by a CsI(Tl) electromagnetic calorimeter. We use Monte Carlo simulations of the \babar\ detector based on GEANT4~\cite{GEANT} to optimize selection criteria and determine selection efficiencies.

\section{\B meson reconstruction}
\label{sec:Breconstruction}

We reconstruct \Bp and \Bz decays (\Brec) in the modes $\Bp\to \Db^{(*)0}\pip$, $\Db^{(*)0}\rho^+$, $\Db^{(*)0}a_1^+$ and $\Bz\to D^{(*)-}\pip$, $D^{(*)-}\rho^+$, $D^{(*)-}a_1^+$. \Dzb candidates are reconstructed in the $\Kp\pim$, $\Kp\pim\piz$, $\Kp\pim\pip\pim$ and $\KS\pip\pim$($\KS\to\pip\pim$) decay channels, while \Dm are reconstructed in the $\Kp\pim\pim$ and $\KS\pim$ modes. \Dstar candidates are reconstructed in the $D^{*-}\to \Dzb\pim$ and $\Dstarzb\to \Dzb\piz$ decay modes. 

The kinematic selection of fully reconstructed \B decays relies on two variables. The first is  $\DeltaE = E^*_B-\sqrt{s}/2$, where $E^*_B$ is the energy of the reconstructed \B candidate in the $\ep\en$ center-of-mass frame and $\sqrt{s}$ is the invariant mass of the initial \epem system. The second is the beam-energy substituted mass, defined by $\mes =\sqrt{ (s/2 + {\bf p}_{i}\cdot{\bf p}_{B})^{2}/E_{i}^{2}-{\bf p}^{2}_{B}}$, where ${\bf p}_{B}$ is the \Brec momentum and $(E_{i},{\bf p}_{i})$ is the four-momentum of the initial \epem system, both measured in the laboratory frame. We require $|\DeltaE|<n\,\sigma_{\Delta E}$, using the resolution $\sigma_{\Delta E}$ measured for each decay mode, with $n=2$ or $3$ depending on the decay mode. If an event contains several \Bp (\Bz) candidates, only the highest-purity \B-decay mode is retained. The purity is defined, for each \B-decay mode separately, as the fraction of signal \B decays with $\mes>5.27~\gevcc$, normalized to the total number of reconstructed \Bp (\Bz) candidates in same interval.

\begin{figure}[!t]
\begin{center}
\includegraphics[width=0.85\linewidth]{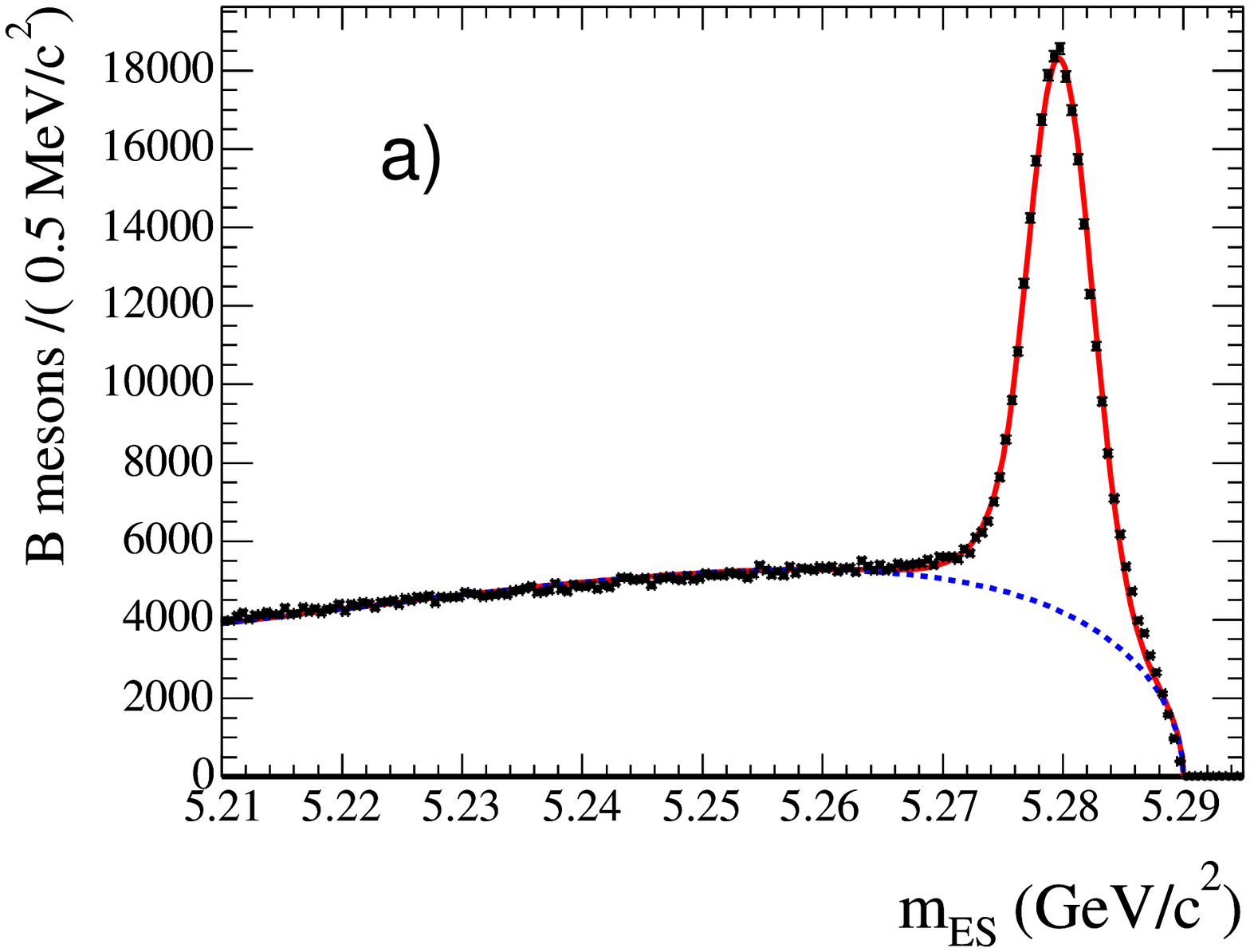}
\includegraphics[width=0.85\linewidth]{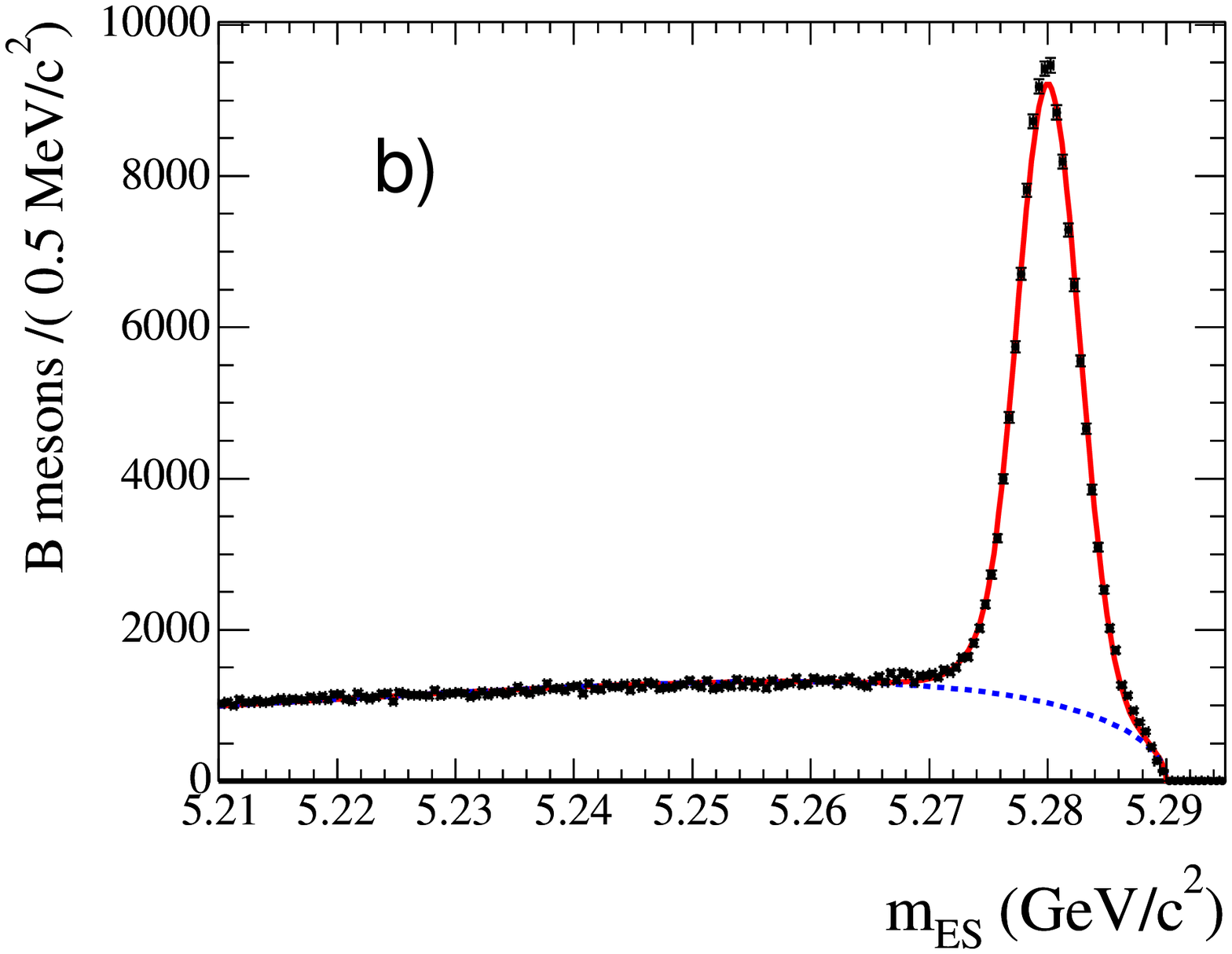}
\caption{\mes spectra of reconstructed (a) \Bp and (b) \Bz candidates. The solid curve is the sum of the fitted signal and background whereas the dashed curve is the background component only.} 
\label{fig:mes}
\end{center}
\end{figure}

\begin{figure}[!t]
    \begin{center}
    \includegraphics[width=1.0\linewidth]{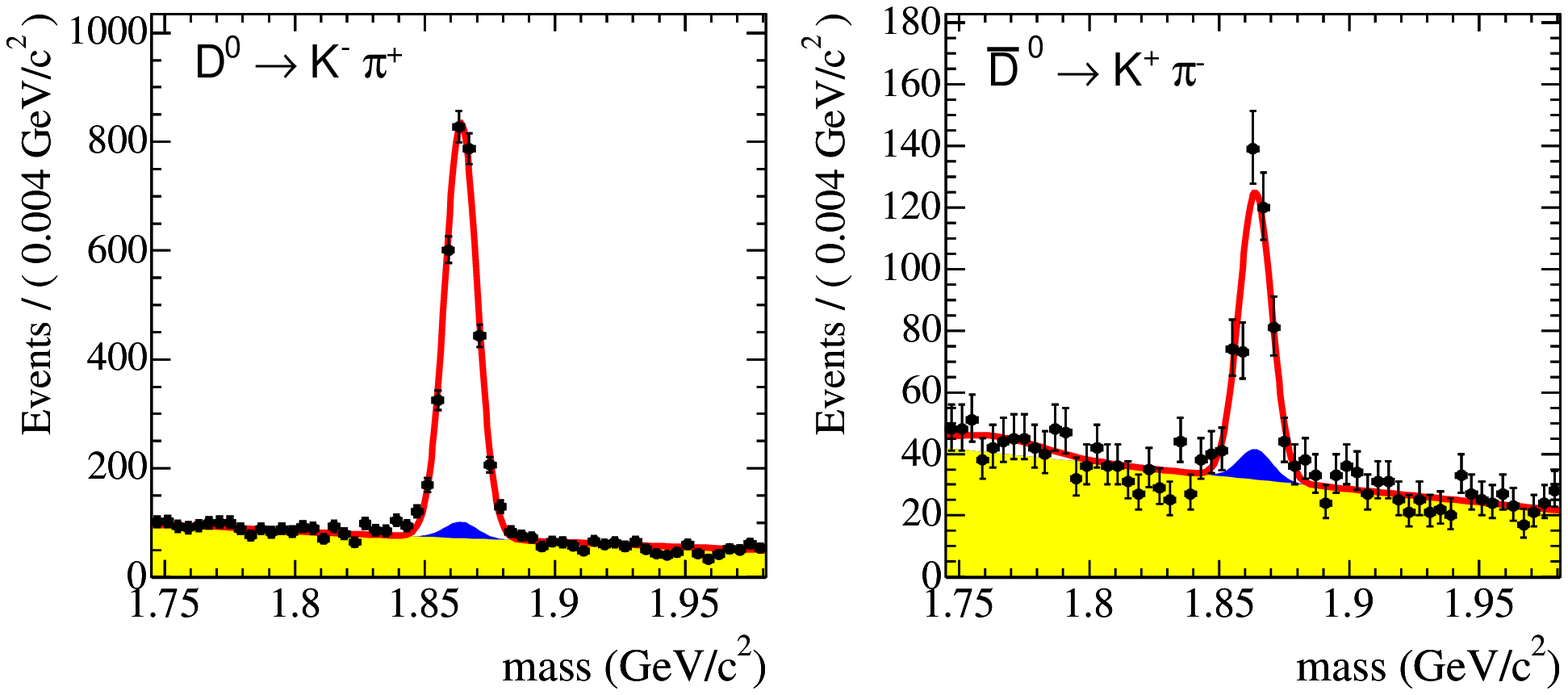}
    \includegraphics[width=1.0\linewidth]{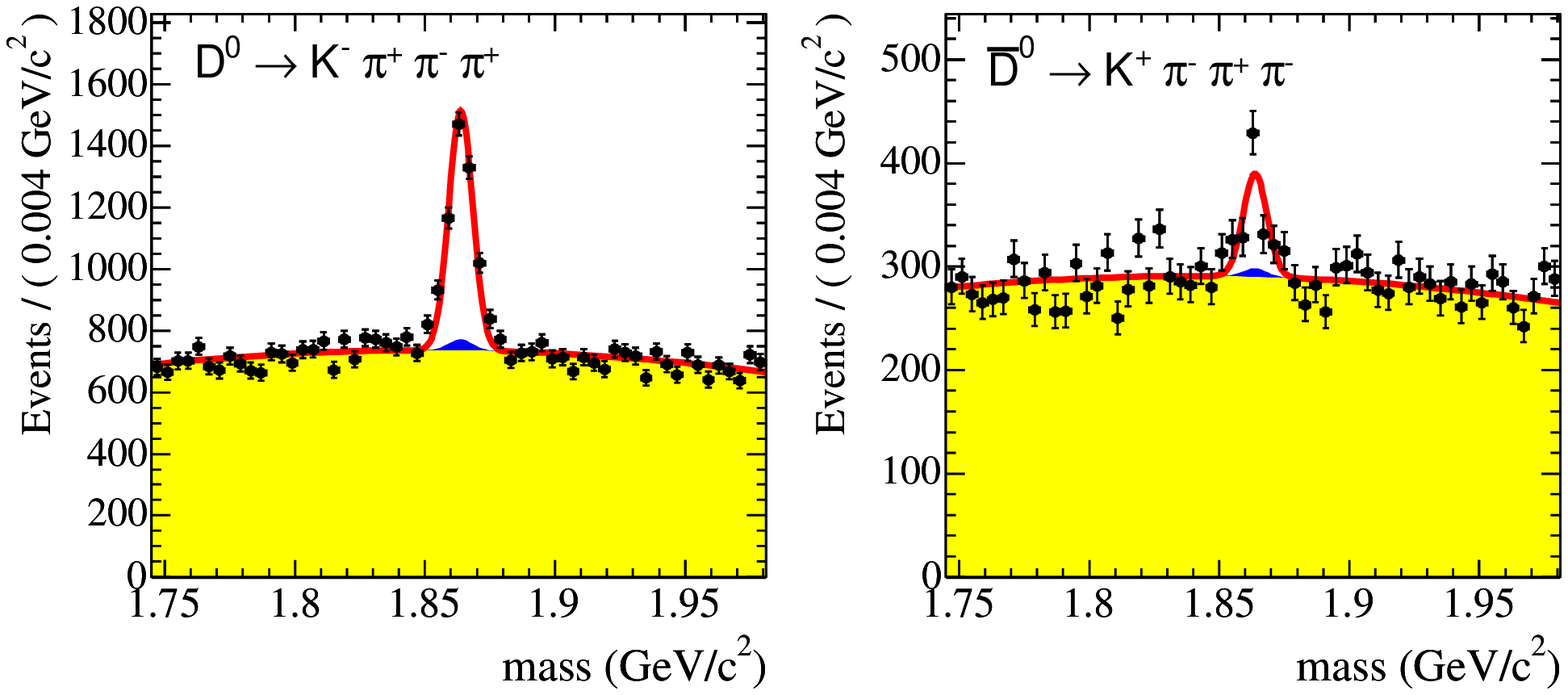}
    \includegraphics[width=1.0\linewidth]{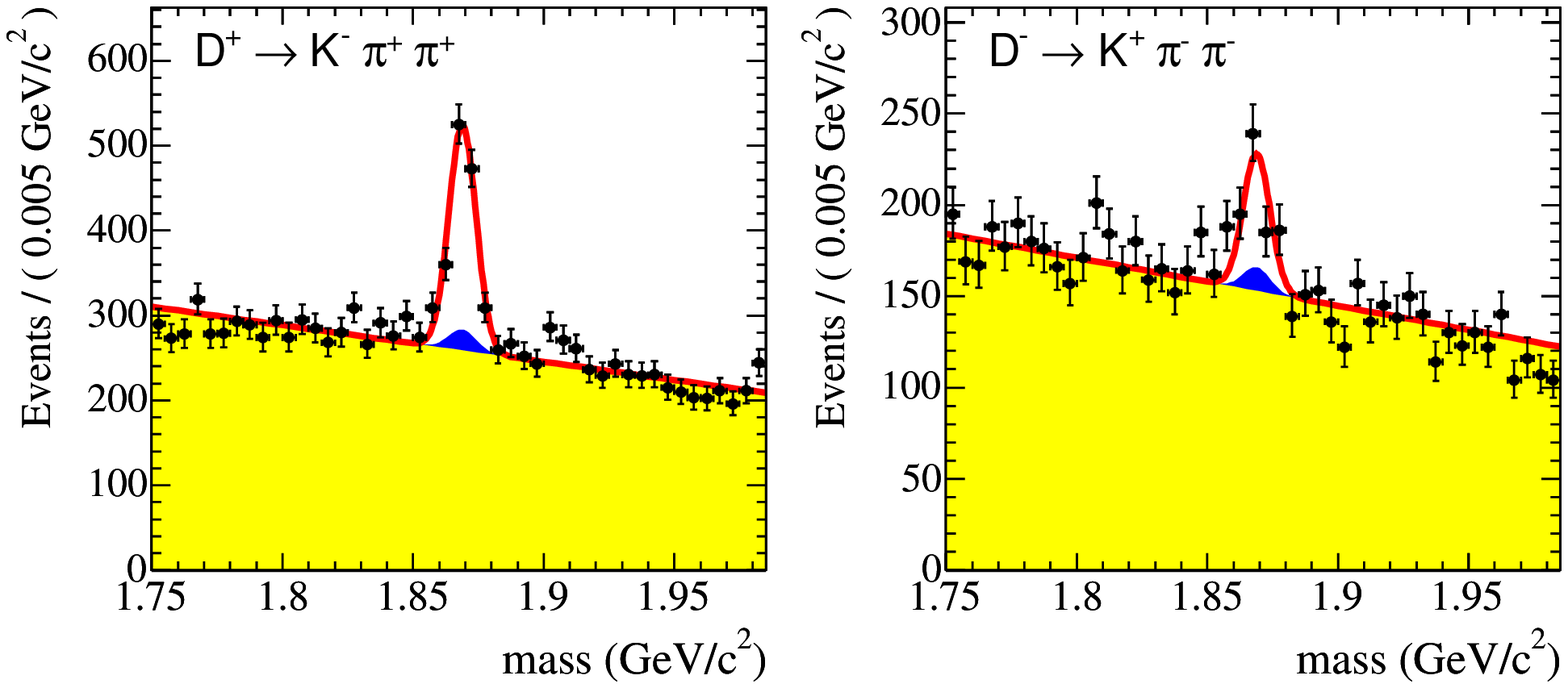}
    \includegraphics[width=1.0\linewidth]{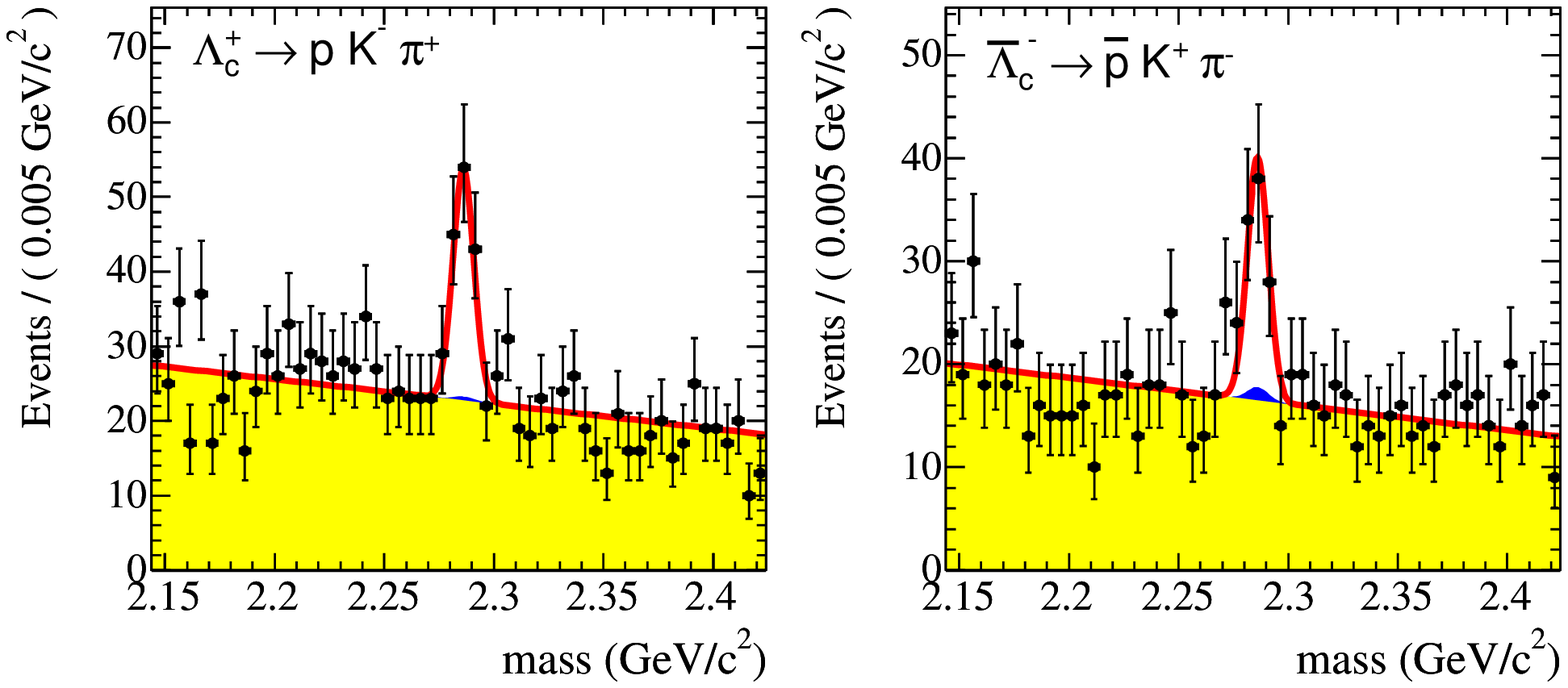}
    \caption{Charm (left) and anti-charm (right) mass spectra in the recoil of \Bp candidates, for the subsample of events with $\mes>5.270~\gevcc$ (\B signal region). The solid curve shows the result of the two-dimensional fit. The dark shaded areas show the contribution of reconstructed $D$,\Db , \Lcp and \Lcm signal in the recoil of combinatorial $\Brec^+$ background. The light shaded area corresponds to the fitted combinatorial (anti-) charm background.}
    \label{fig:dmass_bch}
    \end{center}
\end{figure}
\begin{figure}[!t]
    \begin{center}
    \includegraphics[width=1.0\linewidth]{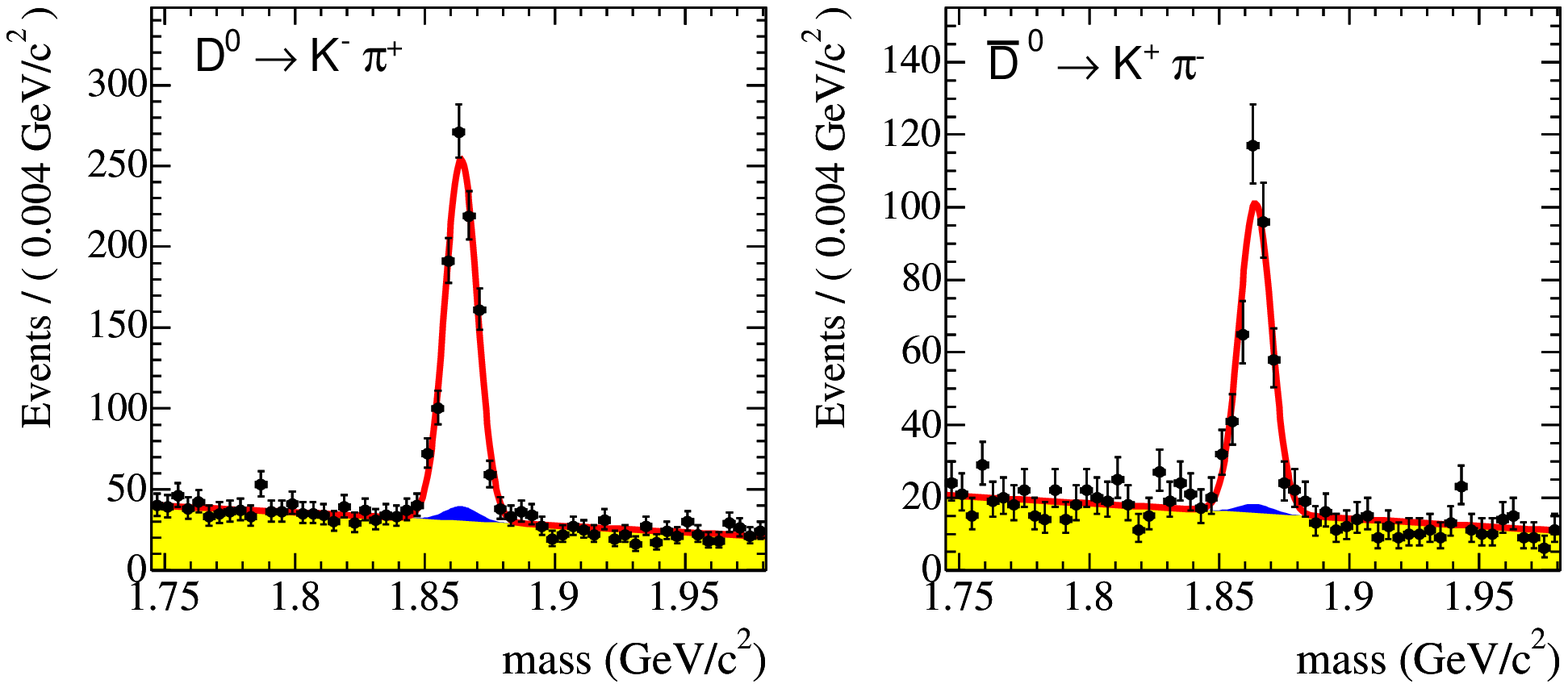}
    \includegraphics[width=1.0\linewidth]{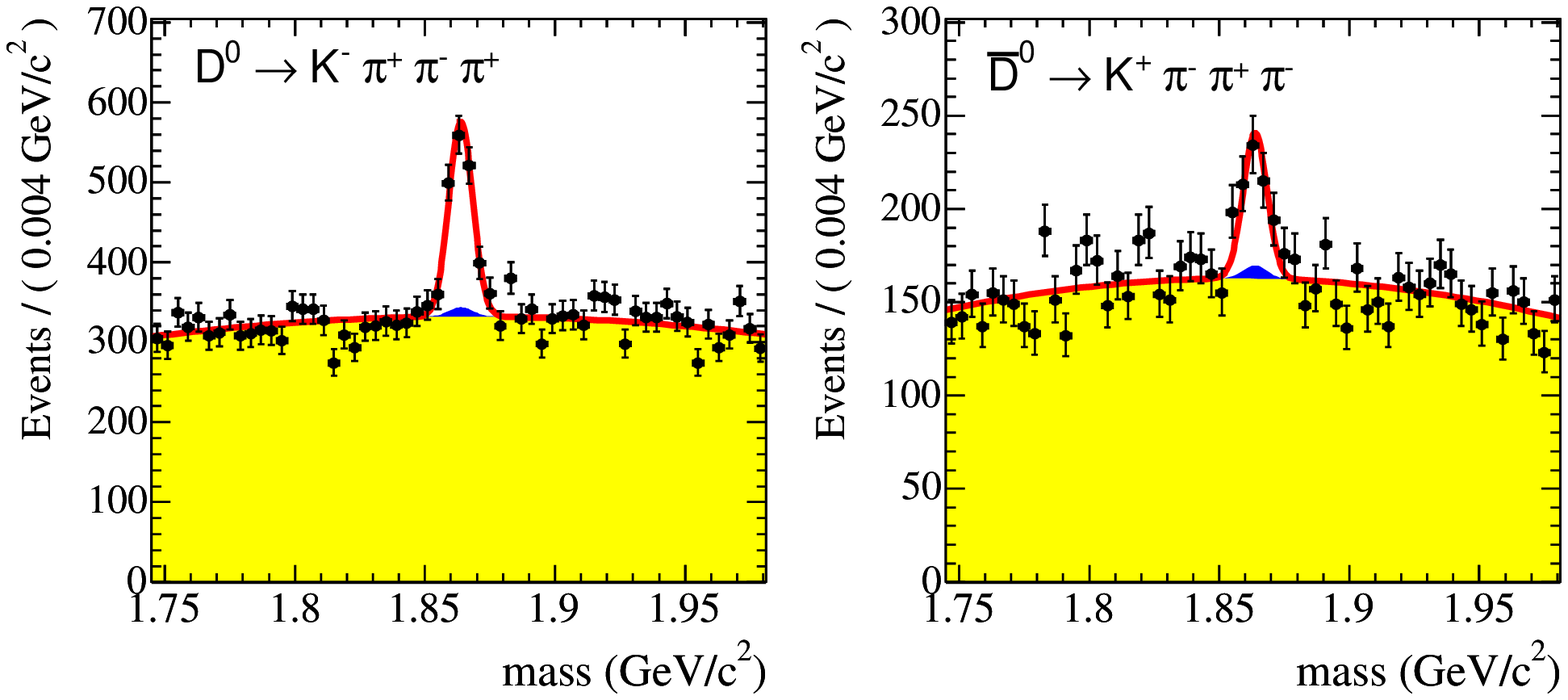}
    \includegraphics[width=1.0\linewidth]{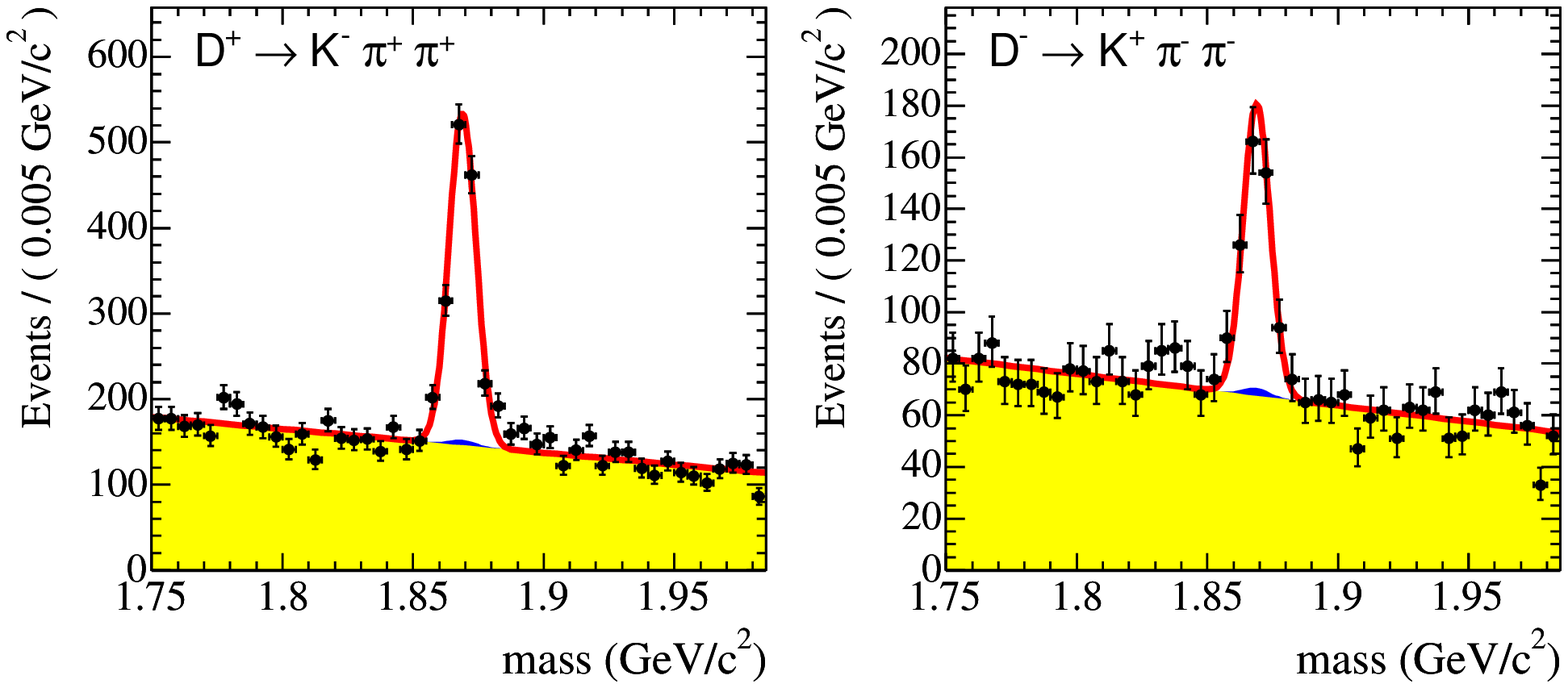}
    \includegraphics[width=1.0\linewidth]{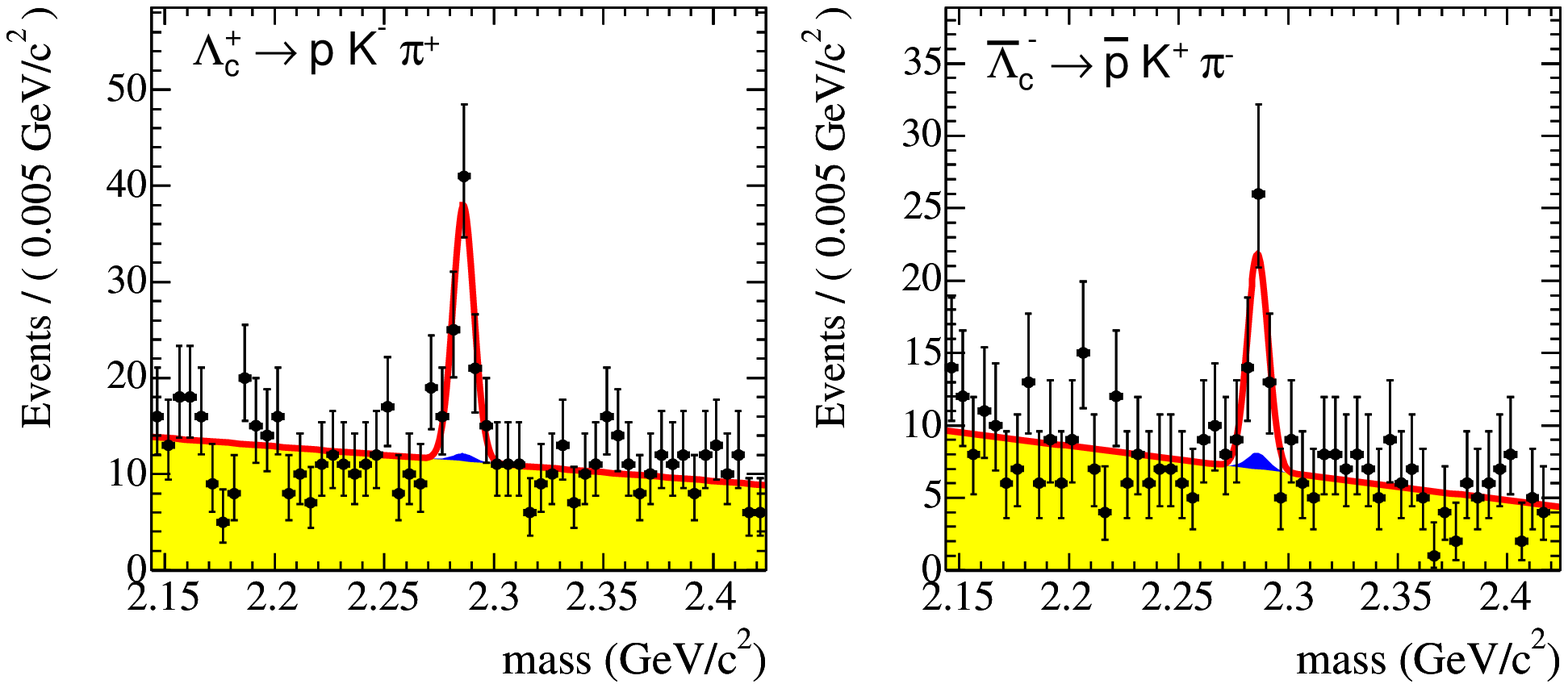}     
	\caption{Charm (left) and anti-charm (right) mass spectra  as for Fig.~\ref{fig:dmass_bch} but in the recoil of \Bzb candidates.}
    \label{fig:dmass_b0}
    \end{center}
\end{figure}
\begin{figure}[!t]
    \begin{center}
    \includegraphics[width=1.0\linewidth]{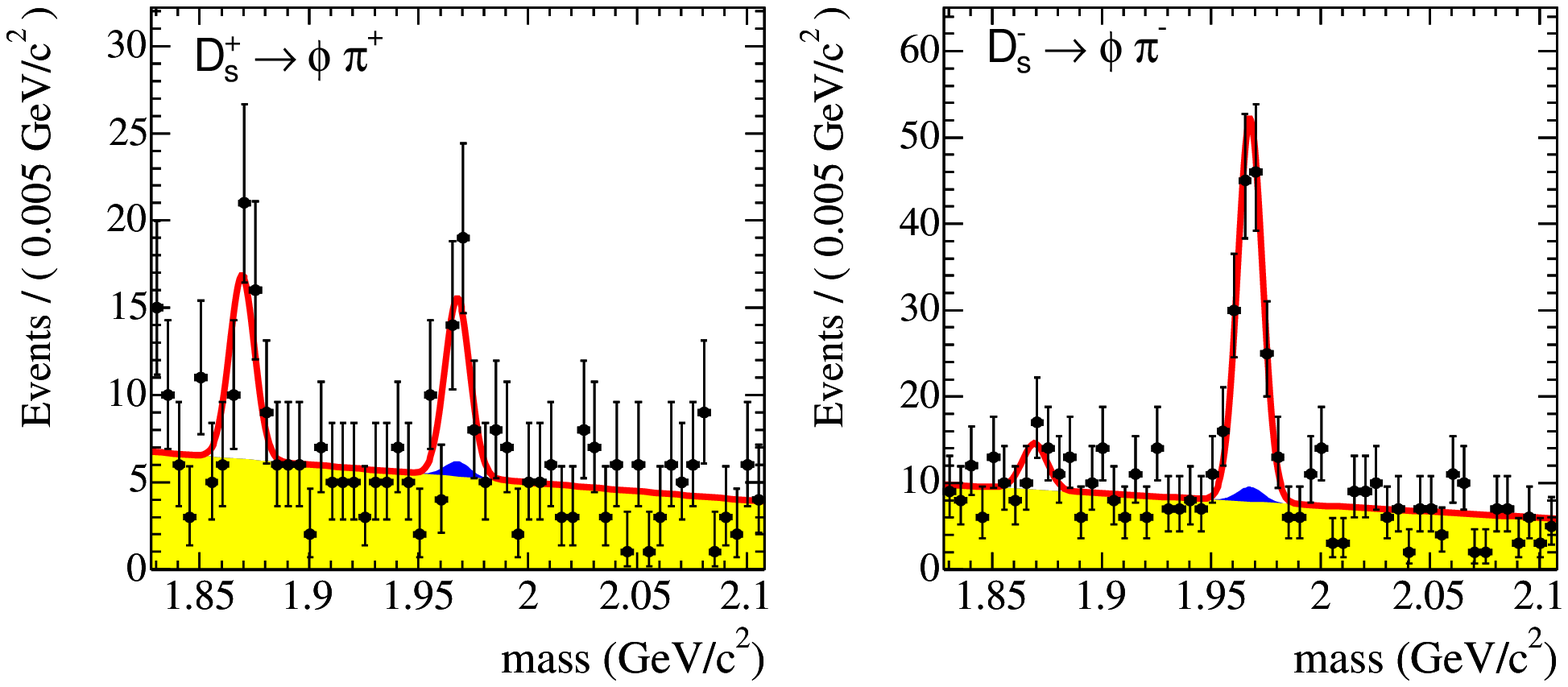}
    \includegraphics[width=1.0\linewidth]{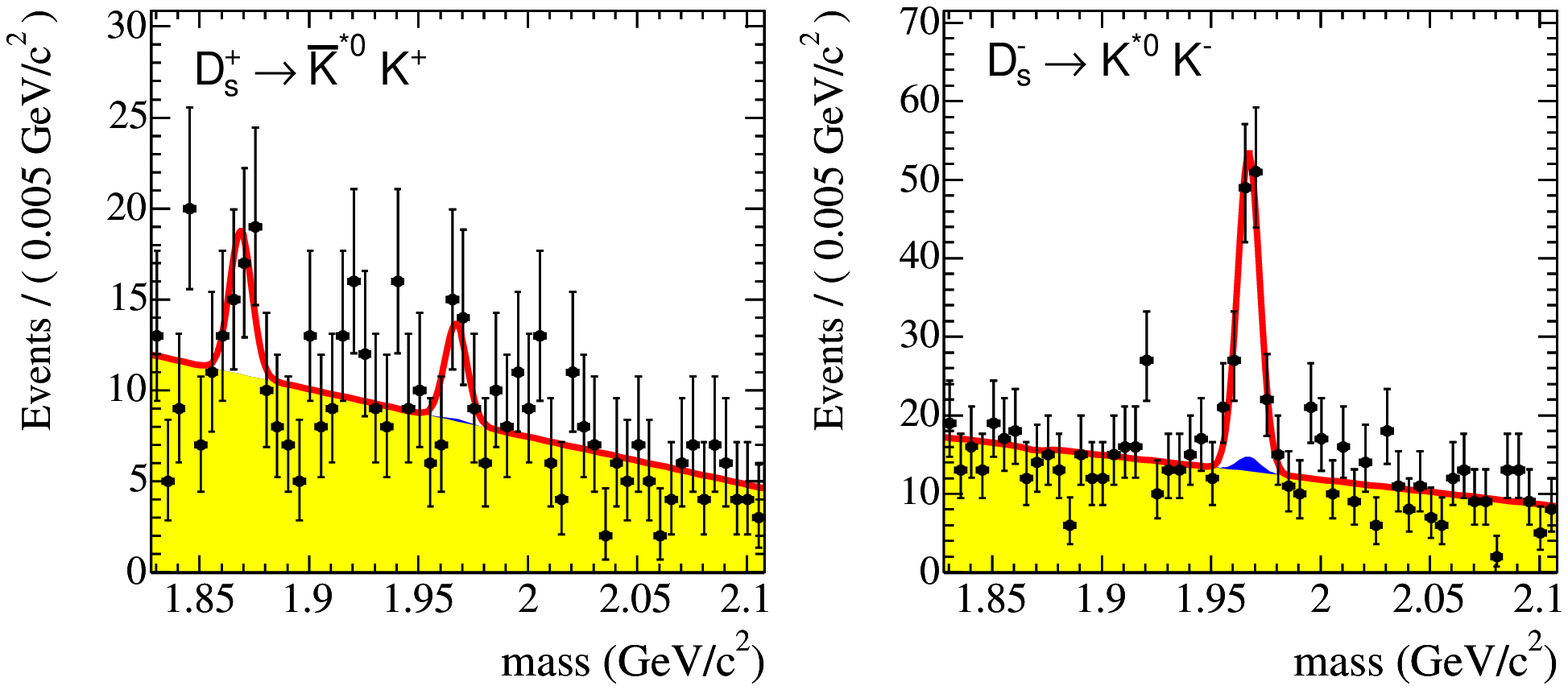}
    \includegraphics[width=1.0\linewidth]{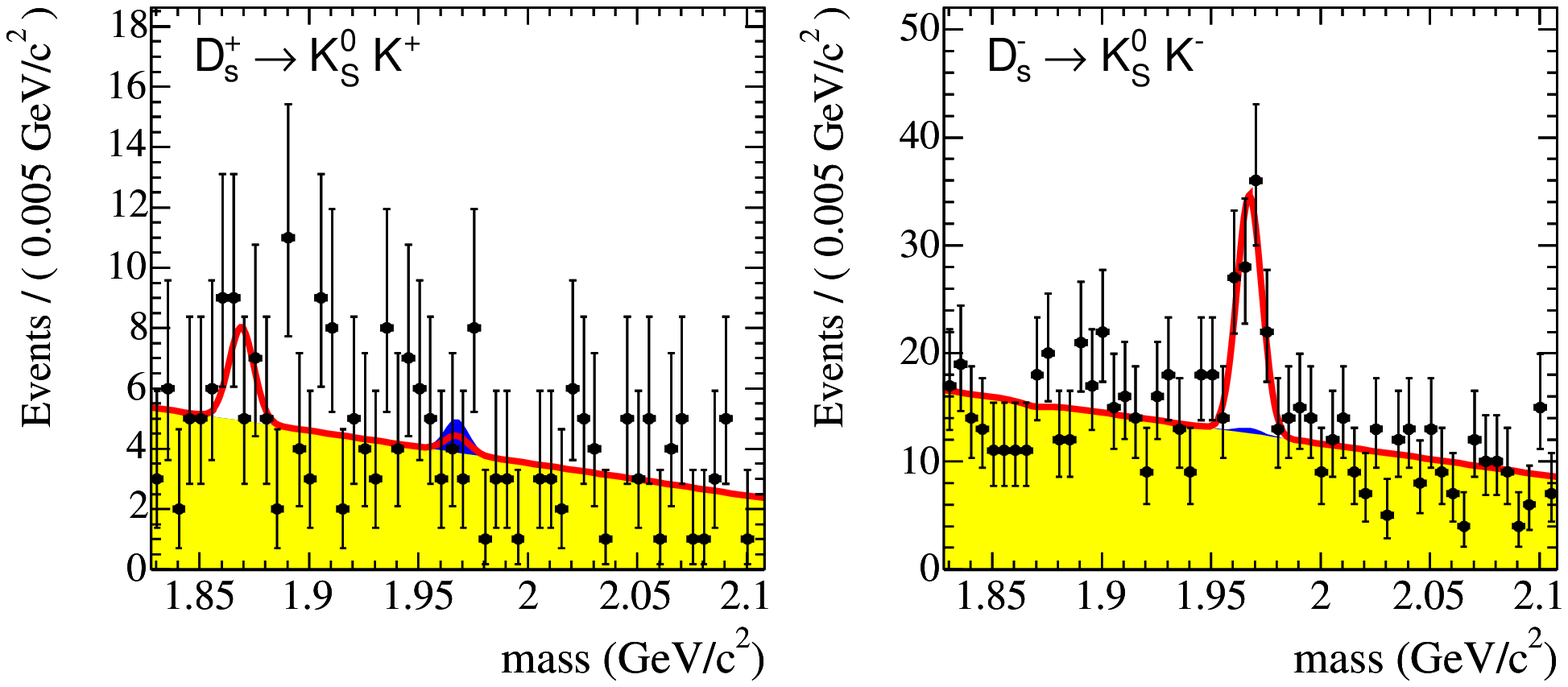}
    \caption{\Dsp (left) and \Dsm (right) mass spectra in the recoil of \Bp candidates, for the subsample of events with $\mes>5.270~\gevcc$ (\B signal region). The solid curve shows the result of the two-dimensional fit. The dark shaded areas show the contribution of reconstructed \Dsp, \Dsm signal in the recoil of combinatorial $\Brec^+$ background. The light shaded area corresponds to the fitted combinatorial (anti-) charm background. The Gaussian peak at the \Dp mass accounts for reconstructed \Dp signal~\cite{satellite_contributions}.}
    \label{fig:dsmass_bch}
    \end{center}
\end{figure}
\begin{figure}[!t]
    \begin{center}
    \includegraphics[width=1.0\linewidth]{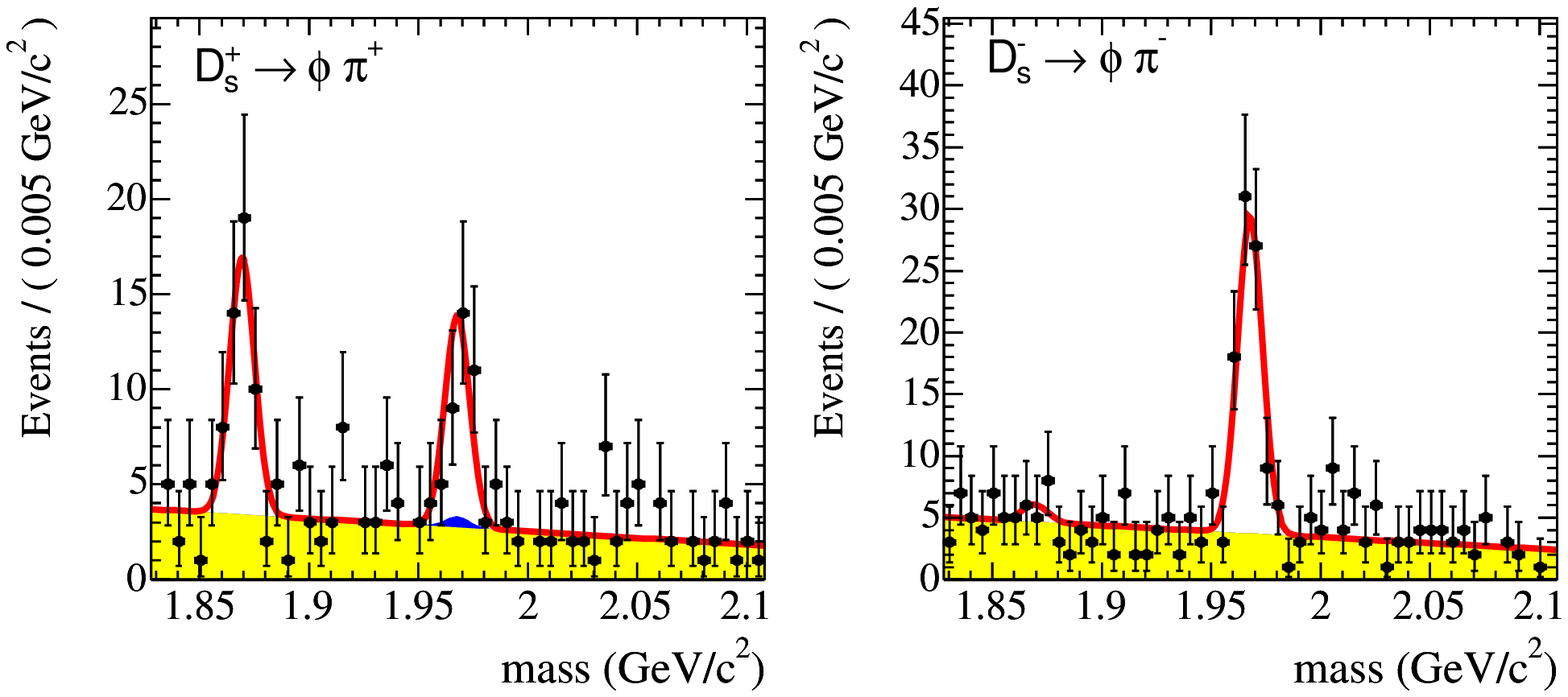}
    \includegraphics[width=1.0\linewidth]{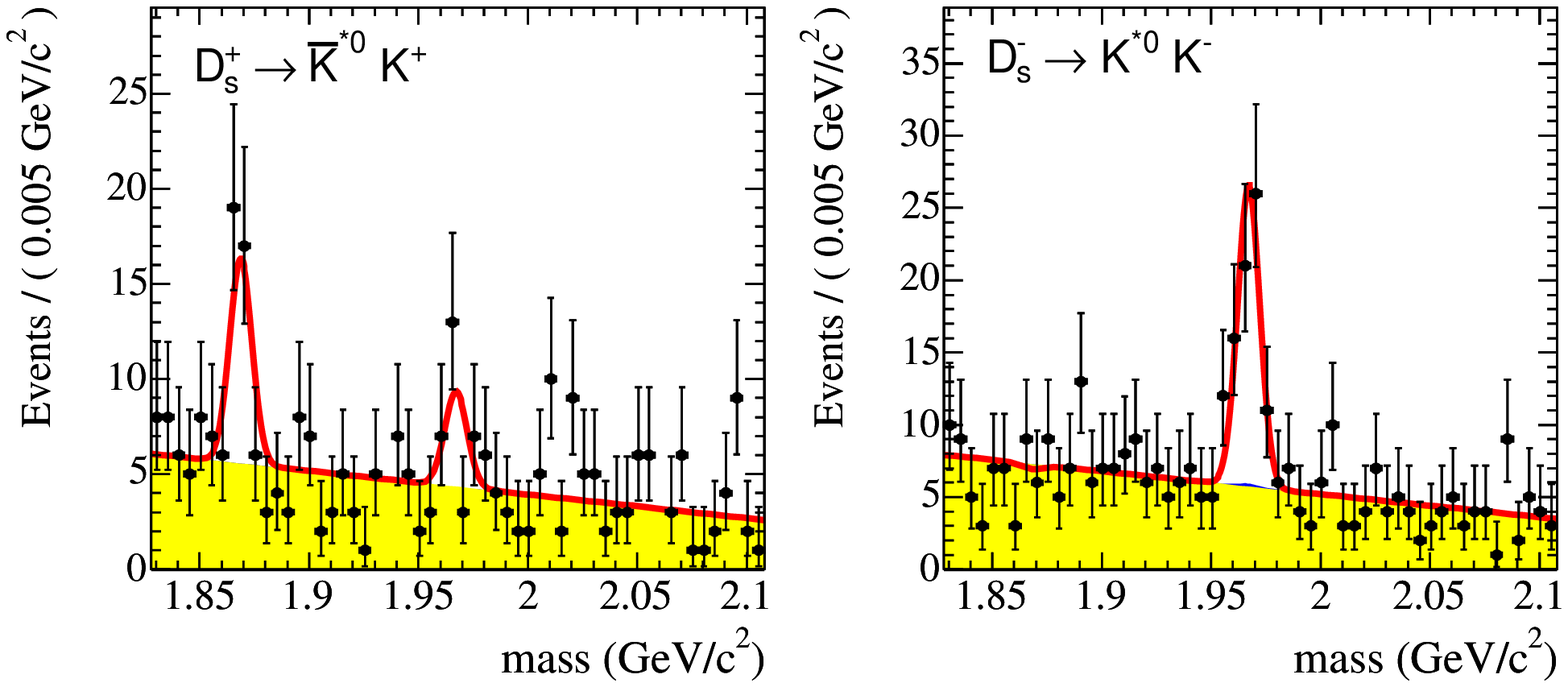}
    \includegraphics[width=1.0\linewidth]{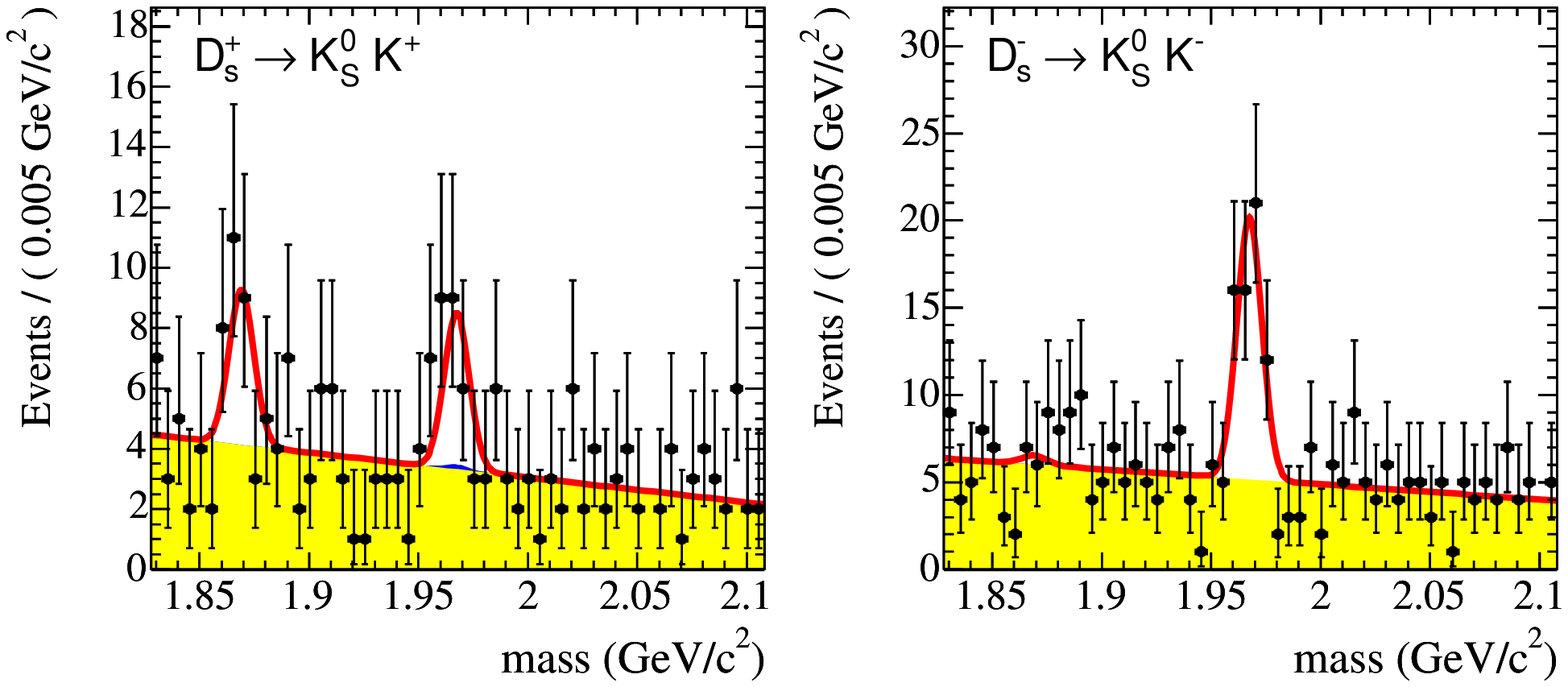}
    \caption{\Dsp and \Dsm mass spectra as for Fig.~\ref{fig:dsmass_bch} but in the recoil of \Bzb candidates.}
    \label{fig:dsmass_b0}
    \end{center}
\end{figure}

The signal yield $N_\B$ of reconstructed \B mesons is extracted from a fit to the \mes spectra (Fig.~\ref{fig:mes}). The \B signal is modeled by a Crystal Ball signal function $\Gamma_{CB}$~\cite{fcrystalball} which is a Gaussian peaking at the \B meson mass modified by an exponential low-mass tail that accounts for photon energy loss. The \B combinatorial background is modeled using the empirical ARGUS phase-space threshold function $\Gamma_{\ARGUS}$~\cite{fargus}. All the signal and background parameters in these functions are extracted from the data. The signal yields of reconstructed \Bp  and \Bz mesons are $N_{\Bp}=200359\pm705$ and $N_{\Bz}=110735\pm424$, where the errors reflect the statistical uncertainty in the number of combinatorial background events. These numbers provide the normalization for all the branching fractions reported below.

The contamination of misreconstructed \Bz events in the \Bp signal (and vice-versa) induces a background which peaks near the \B mass. From the Monte Carlo simulation, the fraction of \Bz events in the reconstructed \Bp signal sample is found to be $c_0=0.038\pm0.009{({\rm syst})}$, and the fraction of \Bp events in the reconstructed \Bz signal sample $c_+=0.028\pm0.007{({\rm syst})}$. The systematic uncertainties take into account possible differences in reconstructing real or simulated events, as well as branching-fraction uncertainties for those \B decay modes contributing to the wrong-charge contamination.

\section{Inclusive charm branching fractions}
\label{sec:CharmCouting}
We now turn to the analysis of inclusive $D$, \Db, \Dsm, \Dsp, \Lcp and \Lcm production in the decays of the $\Bb$ mesons that recoil against the reconstructed \B. Charm particles $\Xc$ are distinguished from anti-charm particles $\Xbc$. They are reconstructed from charged tracks that do not belong to the reconstructed \B. The decay modes considered are listed in Table~\ref{table:decay_modes} along with their branching fractions. Those are taken from Ref.~\cite{PDG2004} except in the case of the $\Dsp\to\phi\pip$ channel~\cite{phipi} for which we use the more precise measurement reported in Ref.~\cite{babarDsPhiPi}.

\begin{table}[hb]
    \begin{center}
    \caption{Charm particle decay modes and branching fractions.}
    \label{table:decay_modes}
    \begin{ruledtabular}
    \begin{tabular}{l l}
$\Xc\to f$             &  $\BR(\Xc\to f)$ (\%)\\ [1mm]
\hline
$\Dz\to\Km\pip$           &  $3.80\pm0.09$ \\[1mm]
$\Dz\to\Km\pip\pim\pip$   &  $7.48\pm0.31$ \\[1mm]
$\Dp\to\Km\pip\pip$          &  $9.1\pm0.7$   \\[1mm]
$\Dsp\to\phi\pip   (\phi\to\Kp\Km)$       & $4.81\pm0.64\,( 49.3 \pm 1.0 \%)$ \\[1mm]
$\Dsp\to\Kstarzb\Kp(\Kstarzb\to\Km\pip)$  & $4.57\pm0.69\,( 66.51\pm 0.01\%)$ \\[1mm] 
$\Dsp\to\KS\Kp     (\KS\to\pip\pim)$      & $2.43\pm0.42\,( 68.95\pm 0.14\%)$ \\[1mm]
$\Lcp\to \proton\Km\pip$          & $5.0\pm1.3$\\
    \end{tabular}
    \end{ruledtabular}
    \end{center}
\end{table}

\subsection{Charm particle yields}
\label{sec:CharmCounting_sec1}
The numbers of charm (anti-charm) particles are extracted from an unbinned maximum likelihood fit to the two-dimensional distribution $[\mes,m_{\Xccbar}]$, where \mes is the beam-energy substituted mass of the reconstructed \B and $m_{\Xccbar}$ is the mass of the charm (anti-charm) particle found among the recoil products. Figs.~\ref{fig:dmass_bch} to~\ref{fig:dsmass_b0} show the results of these fits, projected onto the $m_{\Xccbar}$ axis, for events in the \mes signal region ($\mes>5.270~\gevcc$). The probability density function used to fit the $[\mes,m_{\Xccbar}]$ distributions is the sum of four components~: 

\begin{itemize}
\item {$P_{Bsig}^{Csig}$~:} reconstructed charm (anti-charm) signal in the recoil of reconstructed \B signal,
\item {$P_{Bbkg}^{Csig}$~:} reconstructed charm (anti-charm) signal in the recoil of combinatorial \B background,
\item {$P_{Bsig}^{Cbkg}$~:} combinatorial charm (anti-charm) background in the recoil of reconstructed \B signal,
\item {$P_{Bbkg}^{Cbkg}$~:} combinatorial charm (anti-charm) background in the recoil of combinatorial \B background,
\end{itemize}
These four components are modeled as follows~:
\begin{equation}
\begin{array}{l @{\equiv} r@{\times}l}
P_{Bsig}^{Csig}(\mes,m_{\Xc}) &\ \Gamma_{CB}(\mes)    & \rho_S(m_{\Xc})     \, , \vspace{1.3mm} \\
P_{Bbkg}^{Csig}(\mes,m_{\Xc}) &\ \Gamma_{\ARGUS}(\mes & \rho_S(m_{\Xc})     \, , \vspace{1.3mm} \\
P_{Bsig}^{Cbkg}(\mes,m_{\Xc}) &\ \Gamma_{CB}(\mes)    & \rho_{comb}(m_{\Xc})\, , \vspace{1.3mm} \\ 
P_{Bbkg}^{Cbkg}(\mes,m_{\Xc}) &\ \Gamma_{\ARGUS}(\mes)& \rho_{comb}(m_{\Xc})\, . \vspace{1mm}
\end{array}
\end{equation}
The function~$\Gamma_{CB}$ with all its parameters fixed from the fit detailed in Sec.~\ref{sec:Breconstruction} is used to model the reconstructed \B signal. The combinatorial \B background is described as in Sec.~\ref{sec:Breconstruction} by an ARGUS function $\Gamma_{\ARGUS}$ whose shape parameter is floated in the fit to allow for a possible charm decay-mode dependence of this background. A Gaussian function $\rho_S(m_{\Xccbar})$ describes the mass shape of the reconstructed charm signal. Its mean is taken from the data. Its resolution, as measured in the data, is consistent with that in the simulation and is fixed. The combinatorial charm-background distribution is fitted with a linear function $\rho_{comb}(m_{\Xccbar})$ (except for the \dkpipipi for which a quadratic dependence is assumed)~\cite{satellite_contributions}.\\ 
\begin{table}[!h]
\begin{center}
\caption{$p^*$-averaged reconstruction efficiencies $\epsilon_{\Xc}$ for each charm final state. The errors reflect the limited Monte Carlo statistics.}
\label{tab:effi}
\begin{tabular}{l r}
\hline\hline
$\Xc\to f$  & $\epsilon_{\Xc}$ ($\%$) \\
\hline
\dkpi                  & $50.2\pm0.3$ \\
\dkpipipi \hspace{2mm} & $20.1\pm0.2$ \\
\dkpipi                & $33.7\pm0.2$ \\
\dsphipi               & $33.0\pm0.8$ \\
\dskstark              & $18.0\pm0.5$ \\
\dsksk                 & $31.1\pm0.8$ \\
\lcpkpi                & $26.7\pm0.9$ \\
\hline\hline
\end{tabular}
\end{center}
\end{table}

The reconstruction efficiencies for each charm final state $\Xc\to f$ (Table~\ref{tab:effi}) are computed from the simulation as a function of $p^*$, the charm-particle momentum in the \Bb rest frame, and applied event-by-event to obtain the efficiency-corrected charm and anti-charm signal yields. These are denoted respectively by $N^-(\Xc\to f)$ ($N^0(\Xc\to f)$) and $N^-(\Xbc\to \fbar)$ ($N^0(\Xbc\to \fbar)$) and are listed in Table~\ref{table:t1}. We then determine the charm and anti-charm fractional production rates $\BR_{\c}^{-(0)}$ and $\BRbar_{\c}^{-(0)}$, defined as~:
\begin{equation}
\begin{array}{lcl}
  \BR_\c^{-(0)}           &=& N^{-(0)}(\Xc\to f)      / [N_{\Bp(\Bz)}\times{\BR(\Xc\to f)}]\, , \\
  \BRbar_c^{-(0)} &=& N^{-(0)}(\Xbc\to \fbar) / [N_{\Bp(\Bz)}\times{\BR(\Xc\to f)}]\, ,
\end{array}
\end{equation}
\begin{table*}[!ht]
  \centering
  \caption{Charm and anti-charm efficiency-corrected signal yields and fractional production rates. The uncertainties are statistical only.
}\label{table:t1}
\begin{tabular}{l@{$\to$}l r@{$\pm$}l r@{$\pm$}l r@{$\pm$}l r@{$\pm$}l r@{$\pm$}l r@{$\pm$}l r@{$\pm$}l r@{$\pm$}l}
\hline\hline
  \multicolumn{2}{c}{$\Xc$ decay mode} & 
  \multicolumn{4}{c}{$\Xc$  in recoil of \Brecp} &
  \multicolumn{4}{c}{$\Xbc$ in recoil of \Brecp} & 
  \multicolumn{4}{c}{$\Xc$  in recoil of \Brecz} &
  \multicolumn{4}{c}{$\Xbc$ in recoil of \Brecz}\\
  \multicolumn{2}{c}{  }                & 
  \multicolumn{2}{c}{$N^-(\Xc\to f)$}      
			& \multicolumn{2}{c}{$\BR_c^-$(\%)}&
  \multicolumn{2}{c}{$N^-(\Xbc\to \fbar)$} 
			& \multicolumn{2}{c}{$\BRbar_c^-$(\%)}&
  \multicolumn{2}{c}{$N^0(\Xc\to f)$}      
			& \multicolumn{2}{c}{$\BR_c^0$(\%)}&
  \multicolumn{2}{c}{$N^0(\Xbc\to \fbar)$} 
			& \multicolumn{2}{c}{$\BRbar_c^0$(\%)} \\
  \hline

  $\Dz$  & $\Km\pip$          &$5898$   & $126$ & $77.5$ & $1.6$ & $691$ & $52$  &  $9.1$ & $0.7$  
  			      &$1731$   & $70$  & $41.1$ & $1.7$ & $669$ & $44$  & $15.9$ & $1.0$\\
         & $\Km\pip\pim\pip$  &\hspace{0.5mm} $11010$ & $383$ & $73.4$ & $2.6$ &\hspace{1.5mm} $1378$& $214$ &  $9.2$ & $1.4$
                              &\hspace{1.5mm} $3418$  & $239$ & $41.2$ & $2.9$ &\hspace{1.5mm} $1065$& $159$ & $12.8$ & $1.9$\\
\hline
 $\Dp$  & $\Km\pip\pip$       & $1970$  & $131$ & $10.8$ & $0.7$ & $513$ & $89$  &  $2.8$ & $0.5$
                              &$3044$   & $122$ & $30.2$ & $1.2$ & $869$ & $74$  &  $8.6$ & $0.7$\\
\hline
 $\Dsp$ & $\phi\pip$          &  $85$   & $24$  &  $1.8$ & $0.5$ & $385$ & $42$  &  $8.1$ & $0.9$
                              &  $97$   & $21$  &  $3.7$ & $0.8$ & $227$ & $30$  &  $8.7$ & $1.2$\\
       & $\Kstarzb\Kp$        &  $78$   & $39$  &  $1.3$ & $0.6$ & $567$ & $72$  &  $9.3$ & $1.2$
                              &  $78$   & $28$  &  $2.3$ & $0.8$ & $306$ & $50$  &  $9.1$ & $1.5$\\
          & $\KS \Kp$         &   $0$   & $16$  &  $0.0$ & $0.5$ & $212$ & $39$  &  $6.6$ & $1.2$
                              &  $48$   & $19$  &  $2.7$ & $1.1$ & $148$ & $29$  &  $8.3$ & $1.6$\\
\hline
 $\Lcp$ & $\proton\Km\pip$    & $288$   & $52$  & $2.9$  & $0.5$ & $210$  & $45$ & $2.1$ & $0.5$
                              & $240$   & $41$  & $4.3$  & $0.7$ & $124$  & $30$ & $2.2$ & $0.5$\\
\hline\hline
\end{tabular}
\end{table*} 
where $N_{\Bp}$ ($N_{\Bz}$) is the number of reconstructed \Bp (\Bz) mesons, and $\BR(\Xc\to f)$ is the $\Xc\to f$ branching fraction reported in Table~\ref{table:decay_modes}. $\BR_\c^{-}$, $\BRbar_\c^{-}$, $\BR_\c^{0}$ and $\BRbar_\c^{0}$ are listed in Table~\ref{table:t1}.\\

\subsection{Correlated and anticorrelated charm branching fractions}
\label{sec:CharmCounting_sec2}

\begin{table*}[!th]
\begin{center}
\caption{\Bb branching fractions. The first uncertainty is statistical, the second is systematic, and the third reflects charm branching-fraction uncertainties~\cite{PDG2004,babarDsPhiPi}.}
\label{table:t2}
\begin{tabular}{lcccc}
\hline\hline
                    & \multicolumn{2}{c}{Correlated} & \multicolumn{2}{c}{Anticorrelated}\\[1mm]
$\Xc$\hspace{4mm}  & $\BR(\Bm\to \Xc  \X)$(\%)            & \hspace{2mm} $\BR(\Bzb\to \Xc \X)$(\%)\hspace{2mm} 
                    & \hspace{2mm}$\BR(\Bm\to \Xbc  \X)$(\%) \hspace{2mm} & $\BR(\Bzb\to \Xbc \X)$(\%) \\
\hline \\[-3mm]
$\Dz$   & $78.6\pm1.6\pm2.7^{+2.0}_{-1.9}$ & $47.4\pm2.0\pm1.5^{+1.3}_{-1.2}$
        & $8.6\pm0.6\pm 0.3^{+0.2}_{-0.2}$ & $8.1\pm1.4\pm0.5^{+0.2}_{-0.2}$\\[1mm]
        
$\Dp$   & $9.9 \pm 0.8 \pm 0.5^{+0.8}_{-0.7}$ & $36.9 \pm 1.6 \pm 1.4^{+2.6}_{-2.3}$ 
        & $2.5 \pm 0.5 \pm 0.1^{+0.2}_{-0.2}$ &  $2.3 \pm 1.1 \pm 0.3^{+0.2}_{-0.1}$\\[1mm]
        & &  &                                & $<3.9\rm{\ at\ }90\%\rm{\ CL}$     \\[1mm]
        
$\Dsp$  & $1.1^{+0.4}_{-0.3} \pm 0.1^{+0.2}_{-0.1}$ & $1.5 \pm 0.8 \pm0.1^{+0.2}_{-0.2}$ 
        & $7.9 \pm 0.6 \pm 0.4^{+1.3}_{-1.0}$       & $10.3\pm 1.2 \pm0.4^{+1.7}_{-1.3}$ \\[1mm]
        & & $<2.6\rm{\ at\ }90\%\rm{\ CL}$ &        &\\[1mm]
        
$\Lcp$  & $2.8 \pm 0.5 \pm 0.3^{+1.0}_{-0.6}$ & $5.0\pm1.0\pm0.5^{+1.8}_{-1.0}$
        & $2.1 \pm 0.5 \pm 0.2^{+0.8}_{-0.4}$ & $1.6\pm0.9\pm0.2^{+0.6}_{-0.3}$ \\[1mm]
        & & &                                 & $<3.1\rm{\ at\ }90\%\rm{\ CL}$ \\[1mm]
        
        \hline\hline
\end{tabular}
\end{center}
\end{table*}

For charged \B, the branching fractions for correlated and anticorrelated $\Xc$ production are given by~:
\begin{equation}
    \label{eq:bch_finalBR}
    \begin{array}{l c r}
    \BR(\Bm\to\Xc\X)   &=& \BR_{\c}^-            - c_0 \BR^0_1\, , \vspace{2mm}\\
    \BR(\Bm\to\Xbc\X) &=& \BRbar_{\c}^- - c_0 \BR^0_2\, .
    \end{array}
\end{equation}   
The correlated (anticorrelated) $\Bm\to\Xc\X$ branching fraction is equal to the charm (anti-charm) fractional production rate $\BR_{\c}^-$ ($\BRbar_{\c}^-$ ) in the recoil of reconstructed $\Bp$ mesons modified by a small correction term $c_0\BR^0_1$ ($c_0\BR^0_2$) that accounts for the \Bz contamination in the reconstructed \Bp sample. The factors $\BR^0_1$ and $\BR^0_2$ depend on the measured $\Bzb\to \Xc \X$ and $\Bz\to \Xc \X$ branching fractions, and on the $\BzBzb$ mixing parameter $\chi_d$~\cite{PDG2004}. Doubly Cabibbo-suppressed \Dz decays ($\Dz\to\Kp\pim$ and $\Dz\to\Kp\pip\pim\pim$) are also taken into account. We combine the results from the different \Dz and \Ds decay modes to extract the final branching fractions listed in Table~\ref{table:t2}. The probability of the correlated \Dsp production observed in \Bm decays to be due to a background fluctuation is less than $5\times 10^{-4}$.\\

For neutral \B, charm and anti-charm production in the recoil of reconstructed $\Bz$ mesons have to be corrected for \BzBzb mixing to obtain the correlated and anticorrelated charm branching fractions~:
\begin{equation}
    \label{eq:b0_finalBR}
    \begin{array}{lcr}
    \BR(\Bzb\to\Xc\X)  &=& \frac{\displaystyle \BR_{\c}^0-\chi_d\,(\BR_{\c}^0+\BRbar_{\c}^0)}{\displaystyle 1-2\,\chi_d} 
                           - c_+ \BR^+_1\, ,\vspace{2mm}\\
    \BR(\Bzb\to\Xbc\X) &=& \frac{\displaystyle \BRbar_{\c}^0-\chi_d\,(\BRbar_{\c}^0+\BR_{\c}^0)}{\displaystyle 1-2\,\chi_d} 
	                   - c_+ \BR^+_2\, .
    \end{array}
\end{equation}
The correction factors $c_+ \BR^+_1$ and $c_+ \BR^+_2$ account for \Bp contamination in the \Bz sample and depend on the $\Bm\to \Xc\X$ and $\Bp\to\Xc\X$ branching fractions. Combining the different \Dz and \Ds modes, we obtain the final branching fractions listed in Table~\ref{table:t2}.\\

We also compute the fraction of anticorrelated charm production in \Bb decays~:
\begin{equation} 
\label{eq:w_ac}
w(\Xbc)= \frac{\BR(\Bb\to \Xbc X)}{\BR(\Bb\to \Xc \X)+\BR(\Bb\to \Xbc \X)}.
\end{equation} 
Here, many systematic uncertainties cancel out (tracking, \kaon identification, $D$ branching fractions, \B counting). The results are given in Table~\ref{table:t3}.\\
\begin{table}[!htb]
\begin{center}
\caption{Fraction of anticorrelated charm as defined in Eq.~(\ref{eq:w_ac}).} \label{table:t3}
\begin{ruledtabular}
\begin{tabular}{lcc}
Mode  & $\Bm$ decays & $\Bzb$ decays \\
\hline \\[-3mm]
$\Dzb X$  & $0.098\pm0.007\pm0.001$ & $0.146\pm0.022\pm0.006$ \\[1mm]
$\Dm X$   & $0.204\pm0.035\pm0.001$ & $0.058\pm0.028\pm0.006$ \\[1mm]
          &                         & $<0.098\rm{\ at\ }90\%\rm{\ CL}$ \\[1mm]
$\Dsm X$  & $0.884\pm0.038\pm0.002$ & $0.879\pm0.066\pm0.005$ \\[1mm]
          &                         & $>0.791\rm{\ at\ }90\%\rm{\ CL}$ \\[1mm]
$\Lcm X$  & $0.427\pm0.071\pm0.001$ & $0.243^{+0.119}_{-0.121}\pm0.003$ \\
          &                         & $<0.403\rm{\ at\ }90\%\rm{\ CL}$ \\[1mm]
\end{tabular}
\end{ruledtabular}
\end{center}
\end{table}

The main systematic uncertainties are associated with the track-finding efficiency, the models used to describe the \mes and $m_{\Xccbar}$ distributions, and the particle identification efficiency.  For example, the $2.7\%$ absolute systematic uncertainty on $\BR(\Bm \to \Dz\X)$ reflects the quadratic sum of 
$1.3\%$ attributed to the track-finding efficiency, 
$1.6\%$ to the description of the \mes distribution by the $\Gamma_{\ARGUS}$ and $\Gamma_{CB}$ functions, 
$0.8\%$ to the description of the $m_{\Xccbar}$ signal distribution by the $\rho_{S}$ function,
$1.4\%$ to the particle identification,
$0.5\%$ to the Monte Carlo statistics,  
$0.3\%$ to $c_0$, and $0.1\%$ to $\BR^0_1$. 
 
The uncertainty affecting the track-finding efficiency is estimated with two different methods. The first uses a large inclusive sample of tracks with a minimum number of hits in the silicon vertex detector. The second relies on an $\ep\en\to\taup\taum$ control sample. From these, we derive a relative systematic uncertainty of $0.8\%$ per track. 

The modeling of the \mes distribution by the $\Gamma_{CB}$ and the $\Gamma_{\ARGUS}$ functions affects both the charm signal yields and the numbers of reconstructed \B mesons used in normalizing the branching fractions. The corresponding uncertainty is dominated by the dependence of the $\Gamma_{\ARGUS}$ shape parameter on the lower edge of the \mes fit range. Varying the latter from $5.195$ to $5.225$~\gevcc yields a variation in the branching fraction that is taken as systematic uncertainty. This range was chosen such that the branching fractions measured in the simulation change by $\pm 1$ standard deviation.

The uncertainty associated with the description of the charm signal mass shape by the $\rho_{S}$ function translates into an uncertainty on the charm reconstruction efficiency. It is estimated by fitting the simulated charm signal with a double instead of a single Gaussian.

The systematic uncertainties affecting the proton and charged kaon particle-identification efficiency are estimated using $\Dz\to\Km\pip$ and $\Lambda^0\to\proton\pim$ samples recoiling against reconstructed \Bp and \Bz mesons. The \Dz or $\Lambda^0$ signal yields are extracted in a manner similar to that described in Sec.~\ref{sec:CharmCounting_sec1}, both with and without applying the proton or kaon particle-identification requirements. The ratio of these yields on real and simulated samples is proportional to the particle-identification efficiency in the data and the simulation, respectively. The difference between these two efficiencies is then taken as an estimate of the corresponding the systematic uncertainty ($1.7\%$ relative uncertainty per kaon and $1.3\%$ per proton).

The statistical and systematic uncertainties in Table~\ref{table:t2} and Table~\ref{table:t3} are computed separately for each charm decay mode; correlated errors are taken into account when averaging over \Dz and \Ds final states.  

\subsection{Average charm production in \Bb decays}
\label{sec:nc}
To extract $N_\c$ from the results of Table~\ref{table:t2}, we still need to evaluate the $\Bb\to \Xic \X$ and  $\Bb\to (\ccbar) \X$ branching fractions. Because there exists no absolute measurement of the \Xic-decay branching fraction, the absolute rates for correlated \Xic production in \B decays are unknown~\cite{cleoxicr,belleXicLc}. Therefore, following the discussion in Sec.~\ref{introduction}, we assume that $\BR(\Bb\to \Xic \X) = \BR(\Bb\to \Lcm \X) - \BR(\Bb\to \Lcp \Lcm \Kb(\pi))$~\cite{DLLc}. A recent measurement~\cite{belleLcLcK} indicates that $\Bb\to \Lcp \Lcm \Kb$ decays have a branching fraction of the order of $7\times 10^{-4}$, and thus can be neglected by comparison to $N_\c^{-/0}$ (see also~\cite{babarnc}). We take ${\BR(\Bb \to (\ccbar) X)}$ = $(2.3 \pm 0.3)\%$~\cite{nclep,unkcc} and, using Eqs.~(\ref{eq:nc}) and~(\ref{eq:ncbar}), we obtain for charm production in \Bm decays:
\begin{eqnarray}
\nonumber
  N_\c^-      &=& 0.968 \pm 0.019 \pm 0.032^{+0.026}_{-0.022}, \\ \nonumber
  N_{\cbar}^- &=& 0.234 \pm 0.012 \pm 0.008^{+0.016}_{-0.012}, \\ \nonumber
  n_\c^-      &=& 1.202 \pm 0.023 \pm 0.040^{+0.035}_{-0.029}.
\end{eqnarray}
and in \Bzb decays~:
\begin{eqnarray}
\nonumber
  N_\c^0      &=& 0.947 \pm 0.030 \pm 0.028^{+0.035}_{-0.028}, \\ \nonumber
  N_{\cbar}^0 &=& 0.246 \pm 0.024 \pm 0.009^{+0.019}_{-0.014}, \\ \nonumber
  n_\c^0      &=& 1.193 \pm 0.030 \pm 0.034^{+0.044}_{-0.035}.
\end{eqnarray}

The results reported here are consistent~\cite{consist} with, and supersede those of Ref.~\cite{babarnc}. The three-fold increase in integrated luminosity accounts for the substantial reduction in statistical error. The experimental systematic uncertainties have been similarly reduced, primarily through the use of the two-dimensional $[\mes,m_{\Xccbar}]$ fit, which takes correctly into account the correlation between the fitted number of reconstructed \B mesons and the corresponding charm yield.

\subsection{Isospin analysis}
The main source of anticorrelated \Db mesons produced in \Bb decays is $\b\to\c\cbar\s$ transitions. In these processes isospin should be conserved, leading to the expectation that~: $\Gamma(\Bm\to\Dzb\X) = \Gamma(\Bzb\to\Dm\X)$ and $\Gamma(\Bm\to\Dm\X) = \Gamma(\Bzb\to\Dzb\X)$. However, \Db mesons can also arise from \Dstarb mesons, whose decay does not conserve isospin since the $\Dstarzb\to\Dm\pip$ channel is kinematically forbidden. Thus isospin invariance actually requires~:
\begin{equation}
\begin{array}{l c l}
\Gamma_{dir}(\Bm\to\Dzb\X) &=& \Gamma_{dir}(\Bzb\to\Dm \X)\vspace{1.5mm} \\
\Gamma_{dir}(\Bm\to\Dm \X) &=& \Gamma_{dir}(\Bzb\to\Dzb\X)\vspace{1.5mm}\\
\Gamma(\Bm\to\Dstarzb\X) &=& \Gamma(\Bzb\to\Dstarm \X)\vspace{1.5mm}\\
\Gamma(\Bm\to\Dstarm \X) &=& \Gamma(\Bzb\to\Dstarzb\X)
\end{array}
\label{def:eq}
\end{equation}
where $\Gamma_{dir}(\Bb\to\Db\X)$ refers to the partial width of \Bb-meson decays to \Db mesons where the \Db state is {\it not} reached through a \Dstarb cascade decay. Eqs.~(\ref{def:eq}) lead to the following relations involving the measured anticorrelated \Db branching fractions in Table~\ref{table:t2}~:
\begin{eqnarray}
r\ x^*  = \BR(\Bm \to\Dzb\X)    - \BR(\Bzb\to\Dm \X)\rt&&\label{eq:3} \vspace{1mm}\\
r\ x^*  = \BR(\Bzb\to\Dzb\X)\rt - \BR(\Bm\to\Dm \X)    &&\label{eq:4} \vspace{1mm}
\end{eqnarray}
and~:
\begin{equation}
\begin{array}{lcl}
x+x^* &=& \frac{1}{2}  {\left[\BR(\Bm\to\Dzb\X)    +\BR(\Bm\to\Dm\X)\right.}\vspace{2mm}  \\
    \multicolumn{3}{l}{\left.+\BR(\Bzb\to\Dzb\X)\rt+\BR(\Bzb\to\Dm\X)\rt\right]}
\end{array}
\label{eq:5}
\end{equation}
where $\tau_\Bp/\tau_\Bz$ is the ratio of the \Bp to the \Bz lifetime, $r=\BR(\Dstarm\to\Dzb\pim)$, $x=\BR_{dir}(\Bm\to\Dzb+\Dm\X)$ and $x^*=\BR(\Bm\to\Dstarzb+\Dstarm\X)$~\cite{footnote:x_xstar}.
That both Eqs.~(\ref{eq:3}) and~(\ref{eq:4}) must be satisfied is a consequence of isospin invariance. From these two equations, we extract $x^*$ with a chi-squared method, and using in addition Eq.~(\ref{eq:5}) we calculate~:
\begin{eqnarray}
\BR(\Bm\to\Dstarzb+\Dstarm\X) &=& 9.1 \pm 1.5 \pm 0.6\% \vspace{1mm} \nonumber\\
\BR_{dir}(\Bm\to\Dzb+\Dm\X)   &=& 2.1 \pm 1.7 \pm 0.7\% \vspace{1mm}\nonumber \\
			      &<& 4.5\% \rm{\ at\ }90\%\rm{\ CL}\vspace{1mm}\nonumber \\
\frac{\displaystyle \BR_{dir}(\Bb\to\Dzb+\Dm\X)}{\displaystyle\BR(\Bb\to\Dstarzb+\Dstarm\X)} 
	&=& 0.23_{-0.19}^{+0.25}\pm0.09  \nonumber\\
	&<& 0.60 \rm{\ at\ }90\%\rm{\ CL}\nonumber
\end{eqnarray}
Here the first uncertainty is statistical, the second is systematic and includes charm branching-fraction uncertainties, as well as those affecting the values of $\tau_\Bp/\tau_\Bz$ and $\BR(\Dstarm\to\Dzb\pim)$ taken from Ref.~\cite{PDG2004}. The $\chi^2$ of the fit to Eqs.~(\ref{eq:3}) and~(\ref{eq:4}) is $0.01$ for $1$ degree of freedom.

\section{Charm momentum distributions in the \Bb rest frame}

As the four-momentum of the recoiling \Bb is fully determined, each reconstructed charm hadron can be boosted into the rest frame of its parent \Bb, yielding the $p^*$ distribution of the corresponding (anti-charm) charm species in the \Bb frame. The number of \Xccbar candidates, their fractional production rates and the $\Bb\to\Xccbar\X$ branching fractions are then determined in each $p^*$ bin by the same methods as in Sec.~\ref{sec:CharmCouting}, separately for \Bm and \Bzb decays. The systematic uncertainties are assumed to be independent of $p^*$, except for the error associated with the \Bz (\Bp) contamination in the \Bp (\Bz) sample~: the latter is computed bin-by-bin with a relative uncertainty on $c_+$ and $c_0$ increased to $100\%$. 

\begin{figure}[!ht]
    \begin{center}
    \includegraphics[width=1.0\linewidth]{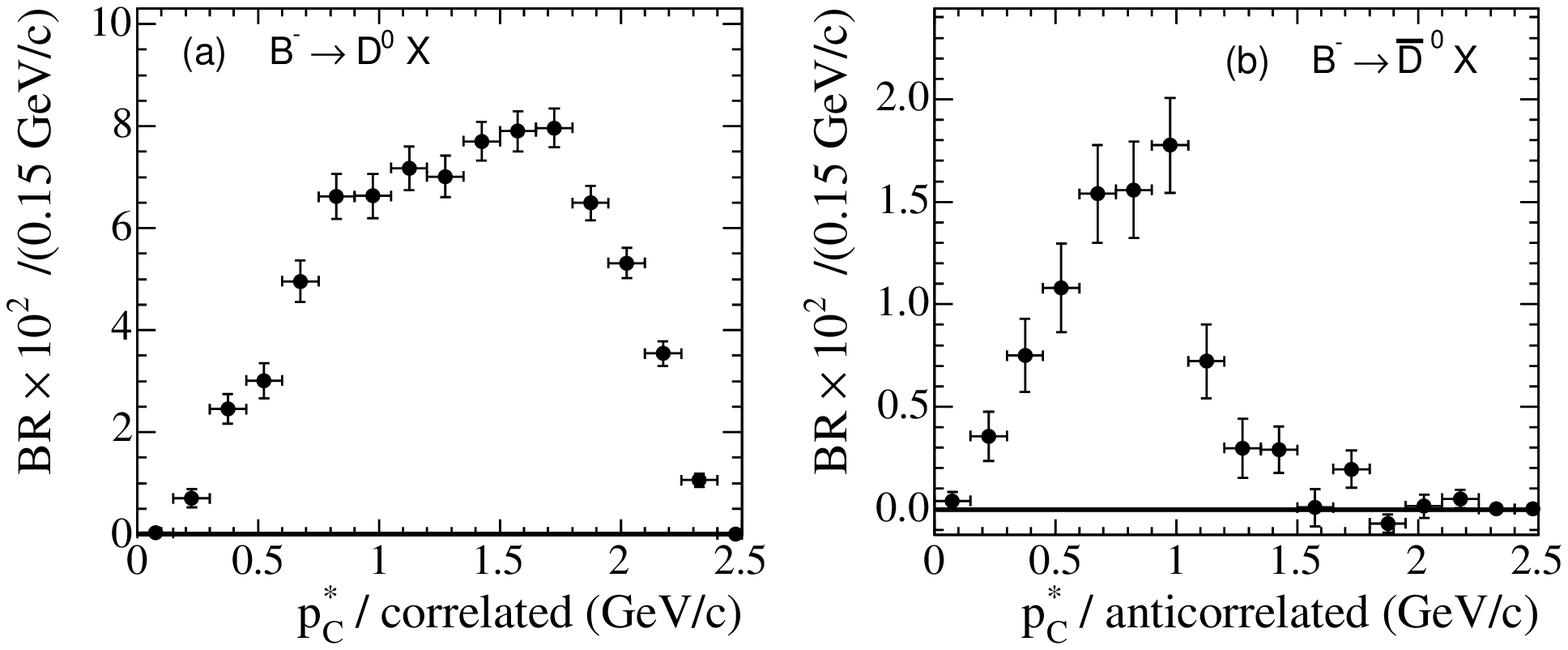}
    \includegraphics[width=1.0\linewidth]{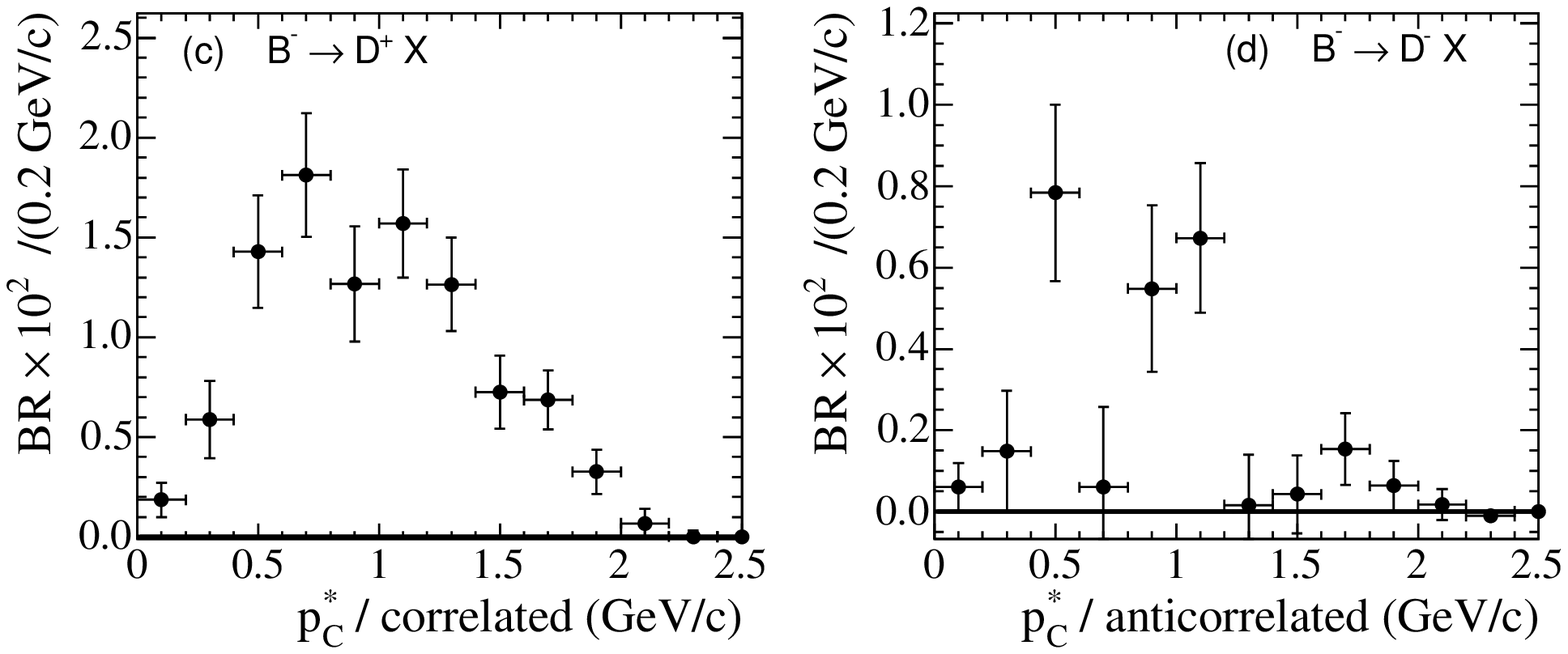}
    \includegraphics[width=1.0\linewidth]{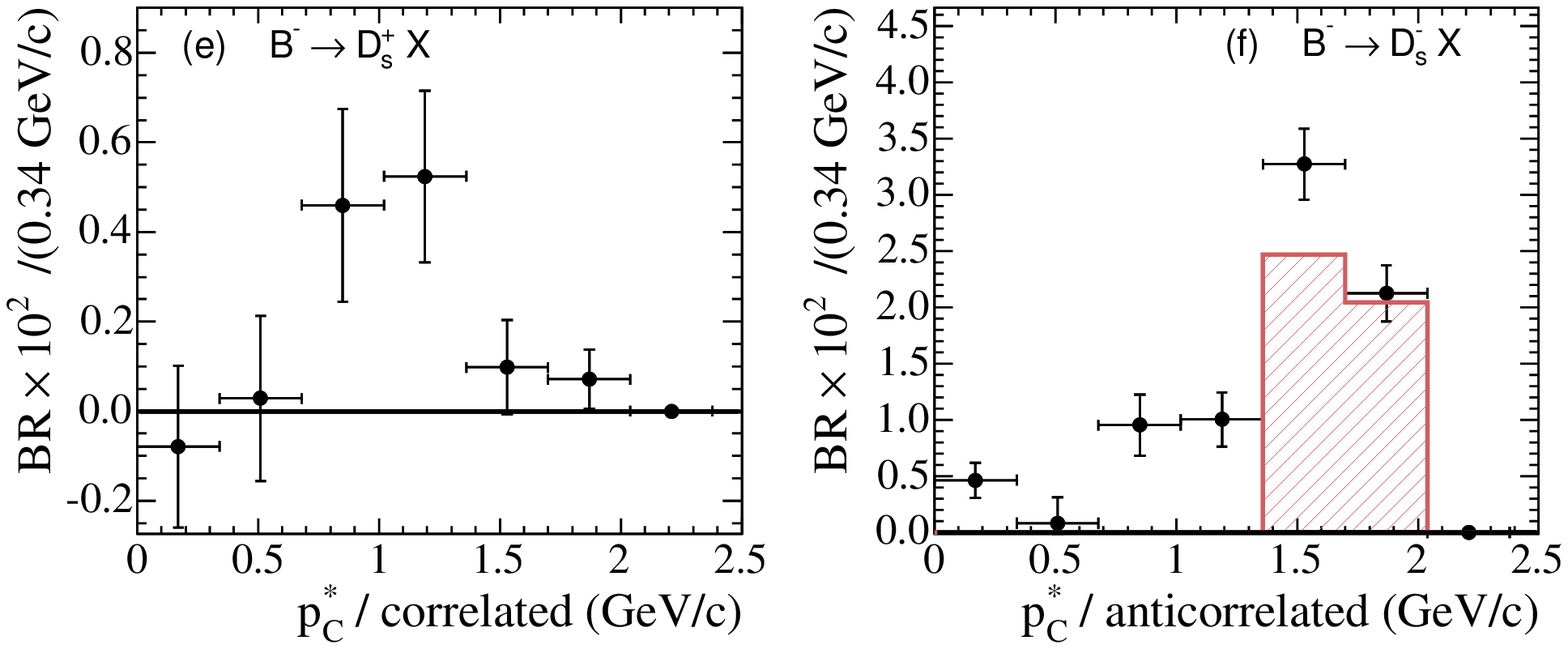}
    \includegraphics[width=1.0\linewidth]{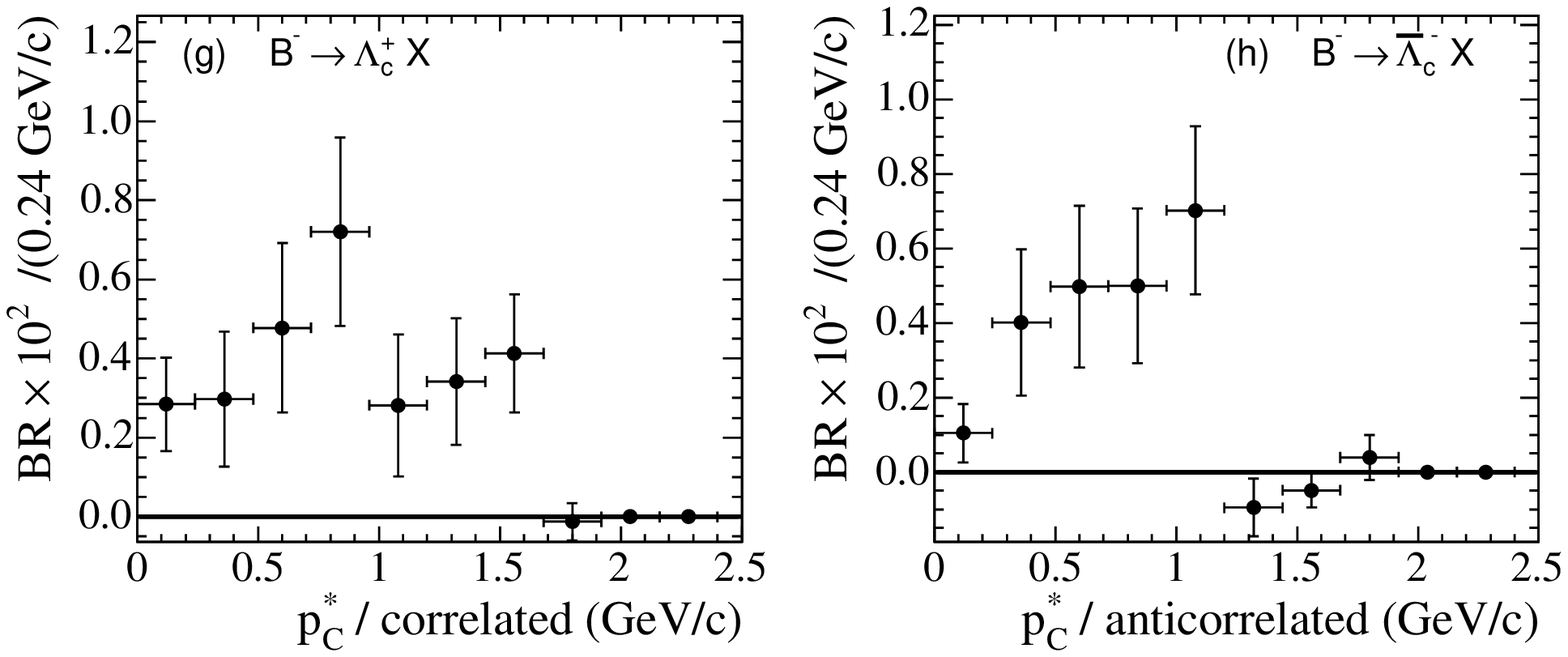}
   \caption{Momentum spectra, in the \Bm rest frame, of correlated (left) and anticorrelated (right) charm particles~: \Dz/\Dzb (a)(b), \Dpm (c)(d), \Dspm (e)(f), \Lcpm (g)(h). The error bars are statistical only. The histogram in frame (f) represents the contribution of $\Bm\to D^{(*)0} \Ds^{(*)-}$ two-body decays assuming the branching fractions of Ref.~\cite{PDG2004} and~\cite{babarDsPhiPi}.}
   \label{fig:pstar_bch}
    \end{center}
\end{figure}

\begin{figure}[!ht]
    \begin{center}    
    \includegraphics[width=1.0\linewidth]{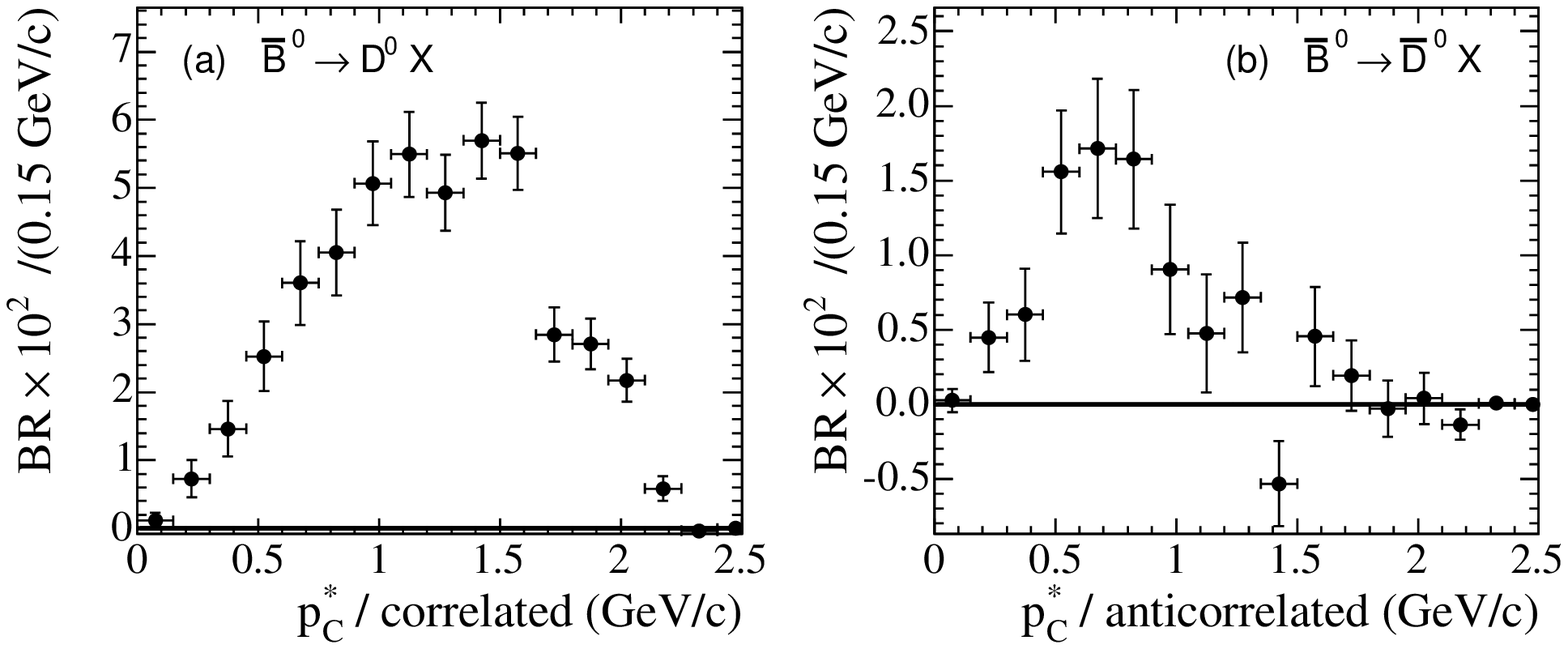}
    \includegraphics[width=1.0\linewidth]{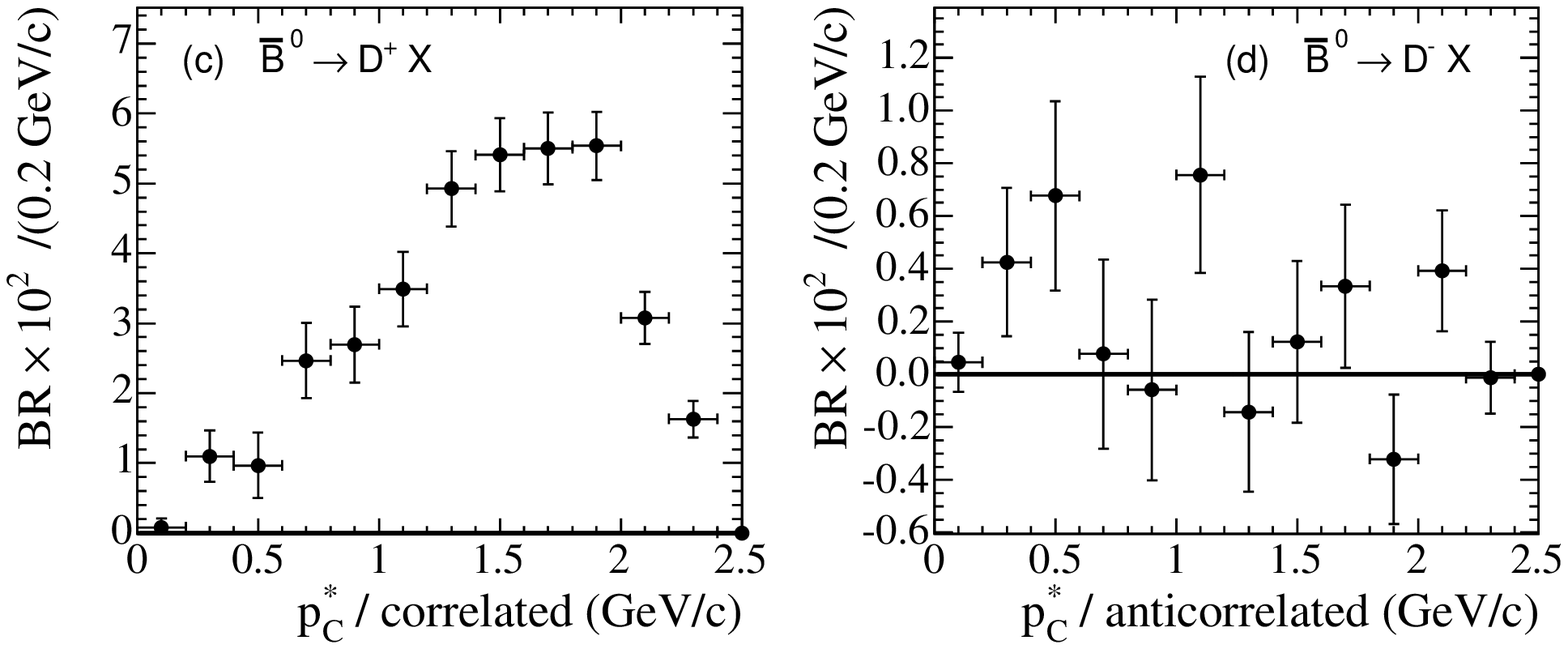}
    \includegraphics[width=1.0\linewidth]{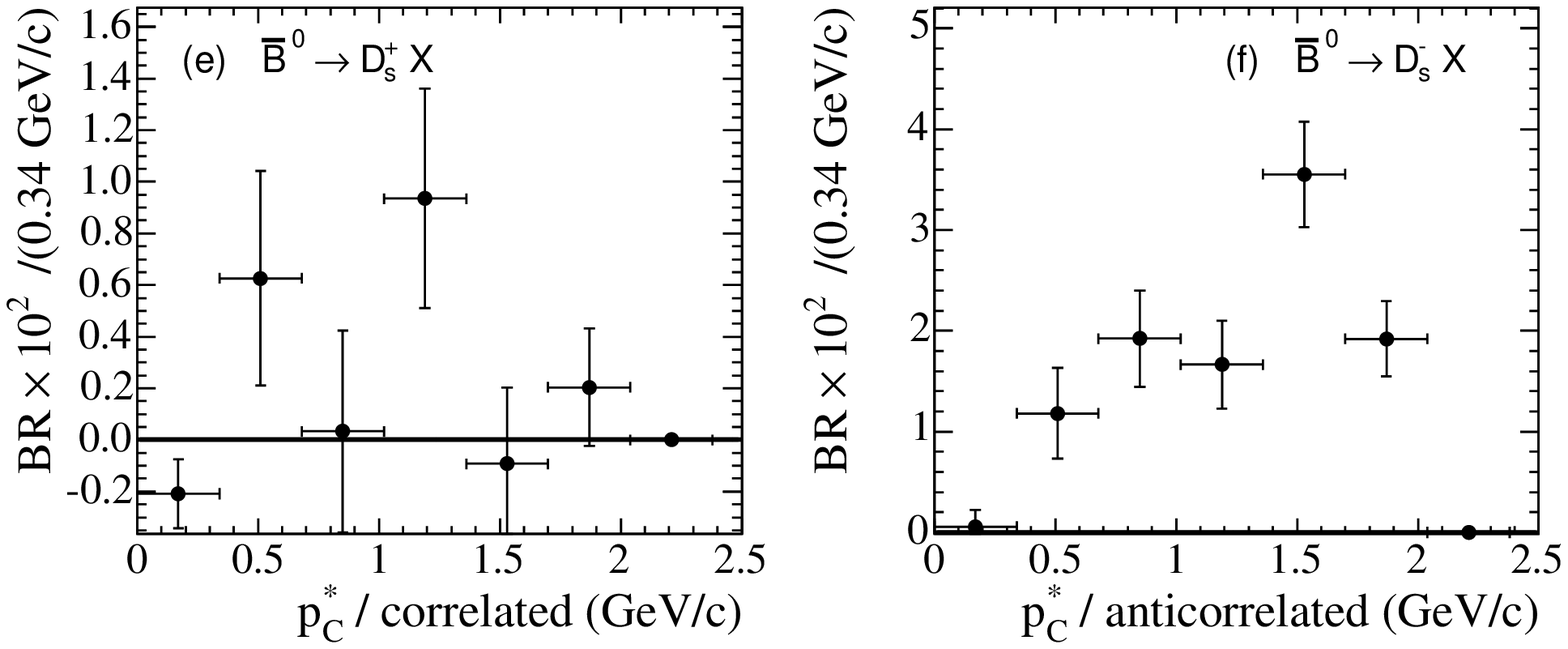}
    \includegraphics[width=1.0\linewidth]{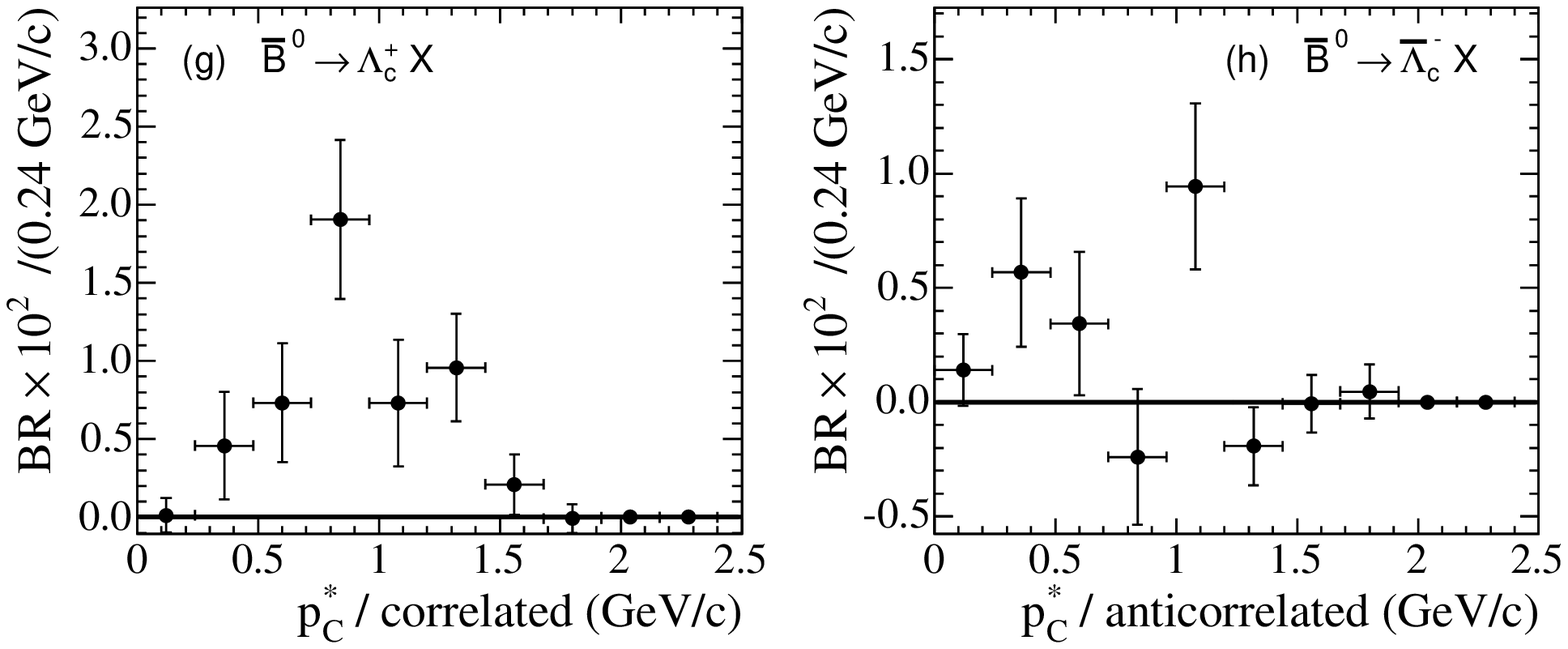}    
    \caption{Momentum spectra, in the \Bzb rest frame, of correlated (left) and anticorrelated (right) charm particles~: \Dz/\Dzb (a)(b), \Dpm (c)(d), \Dspm (e)(f), \Lcpm (g)(h). The error bars are statistical only.} 
    \label{fig:pstar_b0}
    \end{center}
\end{figure}

Figs.~\ref{fig:pstar_bch} and~\ref{fig:pstar_b0} show the result for correlated and anticorrelated \Dz, \Dp, \Ds and \Lc production in \Bm and \Bzb decays, respectively. The numerical values are tabulated in the Appendix.\\

Correlated \Dz and \Dp (Figs.~\ref{fig:pstar_bch}a, c and~\ref{fig:pstar_b0}a, c) are produced in several types of transitions~: $\b\to\c\ell^-\nu$, $\b\to\c\ubar\d$ and $\b\to\c\cbar\s$ which explains the fairly large spread of their momentum. High-$p^*$ correlated $D$'s are produced in two-body decays such as $\Bm\to\Dz\pim$ while low momentum $D$'s might come from higher multiplicity final states such as $\Bb\to D \Db K (\X_{light})$ where $\X_{light}$ is any number of pions and/or photons. The latter processes are also the main source of anticorrelated \Dzb and \Dm production (Figs.~\ref{fig:pstar_bch}b, d and~\ref{fig:pstar_b0}b, d) which explains why anticorrelated \Db spectra are softer than their correlated counterparts.

Anticorrelated \Dsm spectra (Figs.~\ref{fig:pstar_bch}f and~\ref{fig:pstar_b0}f) have a very different shape compared to anticorrelated \Db spectra. They are peaked at high $p^*$ values which is suggestive of the two-body decays $\Bb\to D^{(*)} \Dsm$  and  $\Bb\to D^{(*)} \Ds^{*-}$. These decays represent a large fraction of the total anticorrelated \Dsm production as shown in Fig.~\ref{fig:pstar_bch}. In contrast, the corresponding two-body processes $\Bb\to D^{(*)}\Dm$ and $\Bb\to D^{(*)}\Dstarm$ are Cabibbo-suppressed.

In the case of anticorrelated \Lcm production associated with \Xic production, for decays such as $\Bb\to\Xic\Lcm(\X_{light})$, the anticorrelated \Lcm spectra should have a cut-off at $p^* < 1.15~\gevc$.  This is actually observed in the data, both in \Bm (Fig.~\ref{fig:pstar_bch}h) and in \Bzb (Fig.~\ref{fig:pstar_b0}h) decays.

\section{Conclusions}

We have measured the branching fractions for inclusive decays of $B$ mesons to flavor-tagged $D$, \Ds and \Lc, separately for \Bm and \Bzb. We observe a significant production of anticorrelated \Dz and \Dp mesons in \B decays, with the branching fractions reported in Table~\ref{table:t2}. These results are consistent with and supersede our previous measurement~\cite{babarnc}. We find evidence for correlated \Dsp production in \Bm decays, a process which has not been previously reported.

The sum of all correlated charm branching fractions, $N_\c$, is compatible with 1, for charged as well as for neutral \B mesons. The numbers of charm particles per \Bm decay  ($n_\c^- = 1.202 \pm 0.023 \pm 0.040 ^{+0.035}_{-0.029} $) and per \Bzb decay ($n_\c^0 = 1.193\pm 0.030\pm 0.034^{+0.044}_{-0.035}$) are consistent with previous measurements~\cite{babarnc,cleoinc,nclep} and with theoretical expectations~\cite{ncbagan,ncbuchalla,ncneubert,nc_lenz}.

Assuming isospin conservation in the $\b\to\c\cbar\s$ transition, we show that anticorrelated \Db mesons are mainly produced by cascade decays $\Bb\to\Dstarb\X\to\Db\X$.

Finally, the technique developed for this analysis allows us to measure the inclusive momentum spectra of flavor-tagged $D$, \Ds and \Lc in the rest frame of the \Bb parent, separately in \Bm and \Bzb decays, eventually providing insight into \B-decay mechanisms.\\

\subsection*{Acknowledgments}
We are grateful for the excellent luminosity and machine conditions
provided by our \pep2\ colleagues, 
and for the substantial dedicated effort from
the computing organizations that support \babar.
The collaborating institutions wish to thank 
SLAC for its support and kind hospitality. 
This work is supported by
DOE
and NSF (USA),
NSERC (Canada),
IHEP (China),
CEA and
CNRS-IN2P3
(France),
BMBF and DFG
(Germany),
INFN (Italy),
FOM (The Netherlands),
NFR (Norway),
MIST (Russia), and
PPARC (United Kingdom). 
Individuals have received support from CONACyT (Mexico), 
Marie Curie EIF (European Union),
the A.~P.~Sloan Foundation, 
the Research Corporation,
and the Alexander von Humboldt Foundation.


\appendix

\section*{Appendix~: Charm $p^*$ spectra}
\label{appendix:pstar_bin}

This appendix tabulates the measured $p^*$ dependence of the branching fractions displayed in Figs.~\ref{fig:pstar_bch} and~\ref{fig:pstar_b0}. In Tables~\ref{tab:Pstar_Dz_Bch} to~\ref{tab:Pstar_Lc_B0}, the first uncertainty is statistical, the second is systematic and includes charm branching-fraction uncertainties. Within each table, the statistical uncertainties are uncorrelated whereas the systematic errors are fully correlated.
\clearpage
\begin{table*}[p]
  \centering
  \caption{Correlated and anticorrelated \Dz production in \Bm decays.}
  \begin{tabular}{c r@{$\pm$}c@{$\pm$}l  r@{$\pm$}c@{$\pm$}l} 
  \hline\hline
      & \multicolumn{3}{c}{correlated prod.} & \multicolumn{3}{c}{anticorrelated prod.} \\ 
  $p^{*}$ range (\gevc) & \multicolumn{3}{c}{$\BR(\Bm\to\X_{\c}\X)$ (\%)}   &
                           \multicolumn{3}{c}{$\BR(\Bm\to\X_{\cbar}\X)$ (\%)} \\ 
  \hline
  0.00 - 0.15 	& 0.03&0.06&0.01 	& 0.04&0.04&0.01 	\\ 
  0.15 - 0.30 	& 0.70&0.18&0.03 	& 0.36&0.12&0.02 	\\ 
  0.30 - 0.45 	& 2.45&0.29&0.11 	& 0.75&0.18&0.03 	\\ 
  0.45 - 0.60 	& 3.01&0.34&0.13 	& 1.08&0.22&0.05 	\\ 
  0.60 - 0.75 	& 4.96&0.40&0.22 	& 1.54&0.24&0.07 	\\ 
  0.75 - 0.90 	& 6.62&0.44&0.30 	& 1.56&0.23&0.07 	\\ 
  0.90 - 1.05 	& 6.63&0.43&0.30 	& 1.78&0.23&0.07 	\\ 
  1.05 - 1.20 	& 7.18&0.43&0.32 	& 0.72&0.18&0.04 	\\ 
  1.20 - 1.35 	& 7.01&0.41&0.32 	& 0.30&0.14&0.05 	\\ 
  1.35 - 1.50 	& 7.70&0.38&0.35 	& 0.29&0.11&0.02 	\\ 
  1.50 - 1.65 	& 7.90&0.39&0.36 	& 0.01&0.09&0.05 	\\ 
  1.65 - 1.80 	& 7.96&0.38&0.40 	& 0.20&0.09&0.02 	\\ 
  1.80 - 1.95 	& 6.49&0.33&0.32 	&-0.07&0.04&0.02 	\\ 
  1.95 - 2.10 	& 5.32&0.29&0.26 	& 0.02&0.06&0.02 	\\ 
  2.10 - 2.25 	& 3.54&0.24&0.19 	& 0.05&0.04&0.00 	\\ 
  2.25 - 2.40 	& 1.06&0.13&0.06 	& \multicolumn{3}{c}{-} \\ 
   \hline\hline
  \end{tabular}
  \label{tab:Pstar_Dz_Bch}
\end{table*}

\begin{table*}[p]
  \centering
  \caption{Correlated and anticorrelated \Dp production in \Bm decays.}
  \begin{tabular}{c r@{$\pm$}c@{$\pm$}l  r@{$\pm$}c@{$\pm$}l} 
  \hline\hline
      & \multicolumn{3}{c}{correlated prod.} & \multicolumn{3}{c}{anticorrelated prod.} \\ 
  $p^{*}$ range (\gevc) & \multicolumn{3}{c}{$\BR(\Bm\to\X_{\c}\X)$ (\%)}   &
                           \multicolumn{3}{c}{$\BR(\Bm\to\X_{\cbar}\X)$ (\%)} \\ 
  \hline
  0.00 - 0.20 	& 0.19&0.09&0.02 	& 0.06&0.06&0.01 	\\ 
  0.20 - 0.40 	& 0.59&0.19&0.06 	& 0.15&0.15&0.02 	\\ 
  0.40 - 0.60 	& 1.43&0.28&0.14 	& 0.78&0.22&0.07 	\\ 
  0.60 - 0.80 	& 1.81&0.31&0.17 	& 0.06&0.20&0.02 	\\ 
  0.80 - 1.00 	& 1.27&0.29&0.13 	& 0.55&0.21&0.05 	\\ 
  1.00 - 1.20 	& 1.57&0.27&0.16 	& 0.67&0.18&0.06 	\\ 
  1.20 - 1.40 	& 1.27&0.23&0.16 	& 0.02&0.12&0.03 	\\ 
  1.40 - 1.60 	& 0.72&0.18&0.15 	& 0.04&0.10&0.04 	\\ 
  1.60 - 1.80 	& 0.69&0.15&0.16 	& 0.15&0.09&0.04 	\\ 
  1.80 - 2.00 	& 0.33&0.11&0.16 	& 0.06&0.06&0.03 	\\ 
  2.00 - 2.20 	& 0.07&0.07&0.09 	& 0.02&0.04&0.03 	\\ 
  \hline\hline
  \end{tabular}
  \label{tab:Pstar_Dp_Bch}
\end{table*}

\begin{table*}[p]
  \centering
  \caption{Correlated and anticorrelated \Ds production in \Bm decays.}
  \begin{tabular}{c r@{$\pm$}c@{$\pm$}l  r@{$\pm$}c@{$\pm$}l} 
  \hline\hline
      & \multicolumn{3}{c}{correlated prod.} & \multicolumn{3}{c}{anticorrelated prod.} \\ 
  $p^{*}$ range (\gevc) & \multicolumn{3}{c}{$\BR(\Bm\to\X_{\c}\X)$ (\%)}   &
                           \multicolumn{3}{c}{$\BR(\Bm\to\X_{\cbar}\X)$ (\%)} \\ 
  \hline
  0.00 - 0.34 	&-0.08&0.18&0.02 	& 0.46&0.16&0.07 	\\ 
  0.34 - 0.68 	& 0.03&0.18&0.03 	& 0.08&0.23&0.04 	\\ 
  0.68 - 1.02 	& 0.46&0.22&0.09 	& 0.95&0.27&0.14 	\\ 
  1.02 - 1.36 	& 0.52&0.19&0.11 	& 1.00&0.24&0.15 	\\ 
  1.36 - 1.70 	& 0.10&0.11&0.03 	& 3.27&0.32&0.49 	\\ 
  1.70 - 2.04 	& 0.07&0.07&0.02 	& 2.13&0.25&0.32 	\\ 
  \hline\hline
  \end{tabular}
  \label{tab:Pstar_Ds_Bch}
\end{table*}

\begin{table*}[p]
  \centering
  \caption{Correlated and anticorrelated \Lc production in \Bm decays.}
  \begin{tabular}{c r@{$\pm$}c@{$\pm$}l  r@{$\pm$}c@{$\pm$}l} 
  \hline\hline
      & \multicolumn{3}{c}{correlated prod.} & \multicolumn{3}{c}{anticorrelated prod.} \\ 
  $p^{*}$ range (\gevc) & \multicolumn{3}{c}{$\BR(\Bm\to\X_{\c}\X)$ (\%)}   &
                           \multicolumn{3}{c}{$\BR(\Bm\to\X_{\cbar}\X)$ (\%)} \\ 
  \hline
  0.00 - 0.24 	& 0.28&0.12&0.09 	& 0.10&0.08&0.03 	\\ 
  0.24 - 0.48 	& 0.30&0.17&0.09 	& 0.40&0.20&0.12 	\\ 
  0.48 - 0.72 	& 0.48&0.21&0.15 	& 0.50&0.22&0.15 	\\ 
  0.72 - 0.96 	& 0.72&0.24&0.22 	& 0.50&0.21&0.15 	\\ 
  0.96 - 1.20 	& 0.28&0.18&0.09 	& 0.70&0.23&0.21 	\\ 
  1.20 - 1.44 	& 0.34&0.16&0.11 	&-0.10&0.08&0.03 	\\ 
  1.44 - 1.68 	& 0.41&0.15&0.13 	&-0.05&0.05&0.01 	\\ 
    \hline\hline
  \end{tabular}
  \label{tab:Pstar_Lc_Bch}
\end{table*}

\begin{table*}[p]
  \centering
  \caption{Correlated and anticorrelated \Dz production in \Bzb decays.}
  \begin{tabular}{c r@{$\pm$}c@{$\pm$}l  r@{$\pm$}c@{$\pm$}l} 
  \hline\hline
      & \multicolumn{3}{c}{correlated prod.} & \multicolumn{3}{c}{anticorrelated prod.} \\ 
  $p^{*}$ range (\gevc) & \multicolumn{3}{c}{$\BR(\Bm\to\X_{\c}\X)$ (\%)}   &
                           \multicolumn{3}{c}{$\BR(\Bm\to\X_{\cbar}\X)$ (\%)} \\ 
  \hline
  0.00 - 0.15 	& 0.11&0.12&0.01 	& 0.03&0.08&0.01 	\\ 
  0.15 - 0.30 	& 0.73&0.28&0.03 	& 0.45&0.23&0.03 	\\ 
  0.30 - 0.45 	& 1.46&0.41&0.07 	& 0.60&0.31&0.04 	\\ 
  0.45 - 0.60 	& 2.53&0.51&0.11 	& 1.56&0.41&0.11 	\\ 
  0.60 - 0.75 	& 3.60&0.62&0.16 	& 1.71&0.47&0.12 	\\ 
  0.75 - 0.90 	& 4.05&0.63&0.20 	& 1.64&0.46&0.12 	\\ 
  0.90 - 1.05 	& 5.07&0.61&0.23 	& 0.90&0.43&0.07 	\\ 
  1.05 - 1.20 	& 5.50&0.62&0.25 	& 0.48&0.40&0.06 	\\ 
  1.20 - 1.35 	& 4.93&0.56&0.24 	& 0.72&0.37&0.08 	\\ 
  1.35 - 1.50 	& 5.70&0.56&0.27 	&-0.53&0.29&0.07 	\\ 
  1.50 - 1.65 	& 5.51&0.53&0.27 	& 0.45&0.33&0.09 	\\ 
  1.65 - 1.80 	& 2.85&0.40&0.23 	& 0.19&0.24&0.07 	\\ 
  1.80 - 1.95 	& 2.71&0.37&0.19 	&-0.03&0.19&0.06 	\\ 
  1.95 - 2.10 	& 2.17&0.32&0.16 	& 0.04&0.17&0.05 	\\ 
  2.10 - 2.25 	& 0.58&0.18&0.11 	&-0.14&0.10&0.02 	\\ 
  \hline\hline
  \end{tabular}
  \label{tab:Pstar_Dz_B0}
\end{table*}

\begin{table*}[p]
  \centering
  \caption{Correlated and anticorrelated \Dp production in \Bzb decays.}
  \begin{tabular}{c r@{$\pm$}c@{$\pm$}l  r@{$\pm$}c@{$\pm$}l} 
  \hline\hline
      & \multicolumn{3}{c}{correlated prod.} & \multicolumn{3}{c}{anticorrelated prod.} \\ 
  $p^{*}$ range (\gevc) & \multicolumn{3}{c}{$\BR(\Bm\to\X_{\c}\X)$ (\%)}   &
                           \multicolumn{3}{c}{$\BR(\Bm\to\X_{\cbar}\X)$ (\%)} \\ 
  \hline
  0.00 - 0.20 	& 0.08&0.12&0.01 	& 0.05&0.11&0.01 	\\ 
  0.20 - 0.40 	& 1.10&0.37&0.09 	& 0.42&0.28&0.07 	\\ 
  0.40 - 0.60 	& 0.97&0.47&0.08 	& 0.68&0.36&0.11 	\\ 
  0.60 - 0.80 	& 2.47&0.54&0.19 	& 0.08&0.36&0.02 	\\ 
  0.80 - 1.00 	& 2.70&0.54&0.21 	&-0.06&0.34&0.02 	\\ 
  1.00 - 1.20 	& 3.49&0.53&0.28 	& 0.76&0.37&0.12 	\\ 
  1.20 - 1.40 	& 4.92&0.54&0.39 	&-0.14&0.30&0.04 	\\ 
  1.40 - 1.60 	& 5.41&0.52&0.44 	& 0.12&0.31&0.04 	\\ 
  1.60 - 1.80 	& 5.50&0.51&0.45 	& 0.33&0.31&0.06 	\\ 
  1.80 - 2.00 	& 5.54&0.49&0.45 	&-0.32&0.25&0.06 	\\ 
  2.00 - 2.20 	& 3.08&0.37&0.25 	& 0.39&0.23&0.06 	\\ 
  2.20 - 2.40 	& 1.63&0.26&0.13 	&-0.01&0.14&0.01 	\\ 
  \hline\hline
  \end{tabular}
  \label{tab:Pstar_Dp_B0}
\end{table*}

\begin{table*}[p]
  \centering
  \caption{Correlated and anticorrelated \Ds production in \Bzb decays.}
  \begin{tabular}{c r@{$\pm$}c@{$\pm$}l  r@{$\pm$}c@{$\pm$}l} 
  \hline\hline
      & \multicolumn{3}{c}{correlated prod.} & \multicolumn{3}{c}{anticorrelated prod.} \\ 
  $p^{*}$ range (\gevc) & \multicolumn{3}{c}{$\BR(\Bm\to\X_{\c}\X)$ (\%)}   &
                           \multicolumn{3}{c}{$\BR(\Bm\to\X_{\cbar}\X)$ (\%)} \\ 
  \hline
  0.00 - 0.34 	&-0.21&0.13&0.03 	& 0.06&0.16&0.02 	\\ 
  0.34 - 0.68 	& 0.63&0.42&0.09 	& 1.18&0.45&0.18 	\\ 
  0.68 - 1.02 	& 0.03&0.39&0.01 	& 1.92&0.48&0.29 	\\ 
  1.02 - 1.36 	& 0.94&0.43&0.14 	& 1.66&0.43&0.25 	\\ 
  1.36 - 1.70 	&-0.09&0.29&0.03 	& 3.55&0.52&0.54 	\\ 
  1.70 - 2.04 	& 0.20&0.23&0.04 	& 1.92&0.37&0.29 	\\ 
  \hline\hline
  \end{tabular}
  \label{tab:Pstar_Ds_B0}
\end{table*}

\begin{table*}[p]
  \centering
  \caption{Correlated and anticorrelated \Lc production in \Bzb decays.}
  \begin{tabular}{c r@{$\pm$}c@{$\pm$}l  r@{$\pm$}c@{$\pm$}l} 
  \hline\hline
      & \multicolumn{3}{c}{correlated prod.} & \multicolumn{3}{c}{anticorrelated prod.} \\ 
  $p^{*}$ range (\gevc) & \multicolumn{3}{c}{$\BR(\Bm\to\X_{\c}\X)$ (\%)}   &
                           \multicolumn{3}{c}{$\BR(\Bm\to\X_{\cbar}\X)$ (\%)} \\ 
  \hline
  0.00 - 0.24 	& 0.01&0.11&0.01 	& 0.14&0.16&0.05 	\\ 
  0.24 - 0.48 	& 0.46&0.34&0.15 	& 0.57&0.33&0.19 	\\ 
  0.48 - 0.72 	& 0.73&0.38&0.23 	& 0.34&0.31&0.12 	\\ 
  0.72 - 0.96 	& 1.90&0.51&0.60 	&-0.24&0.30&0.08 	\\ 
  0.96 - 1.20 	& 0.73&0.40&0.23 	& 0.94&0.36&0.32 	\\ 
  1.20 - 1.44 	& 0.96&0.35&0.30 	&-0.19&0.17&0.07 	\\ 
  1.44 - 1.68 	& 0.21&0.19&0.07 	&-0.01&0.13&0.01 	\\ 
  \hline\hline
  \end{tabular}
  \label{tab:Pstar_Lc_B0}
\end{table*}


\begin{thebibliography}{99}
\bibitem{chconj} Charge conjugation is implied for all decay processes mentioned in this paper.
\bibitem{phipi} We consider any $\Kp\Km$ combination with an invariant mass in the range $1010.6<m_{\Kp\Km}<1028.6~\mevcc$ to be a $\phi$ meson when reconstructing the $\Dsp\to\phi(\to\Kp\Km)\pip$ decay.
\bibitem{satellite_contributions} In some cases, a satellite contribution needs to be added. It includes a reflection from $\Dz\to\Km\Kp$ in the $\Dz\to\Km\pip$ mass spectra and a signal at the \Dp mass (from $\Dp\to\phi\pip,\,\Kstarzb\Kp,\,\KS\Kp$ Cabibbo-suppressed decays) in the $\Ds$ mass spectra.
\bibitem{DLLc} We also neglect the contribution of the $\Bb\to D \Lambda^{0} \Lcm$ decays because of the very small phase space available.
\bibitem{footnote:x_xstar} Assuming isospin conservation, we have also~: $x = \BR_{dir}(\Bzb\to\Dzb+\Dm\X)\times\rt$ and $x^* = \BR(\Bzb\to\Dstarzb+\Dstarm\X)\times\rt$
\end{thebibliography}
\end{document}